\documentclass[aps,prd,twocolumn,amsmath,amssymb,showpacs,nofootinbib]{revtex4}

\usepackage{graphicx}% Include figure files
\usepackage{array,color}
\usepackage{mathrsfs}
\usepackage{longtable,lscape}
\usepackage{txfonts}
\usepackage{amssymb}
\usepackage{indentfirst}
\usepackage{color}
\usepackage{amssymb}
\usepackage{epsfig}
\usepackage{multirow}

\newcommand{\be}{\begin{equation}}
\newcommand{\ee}{\end{equation}}
\newcommand{\bea}{\begin{eqnarray}}
\newcommand{\eea}{\end{eqnarray}}
\newcommand{\bean}{\begin{eqnarray*}}
\newcommand{\eean}{\end{eqnarray*}}

\newcommand{\gapproxeq}{\lower
.7ex\hbox{$\;\stackrel{\textstyle >}{\sim}\;$}}
\newcommand{\lapproxeq}{\lower
.7ex\hbox{$\;\stackrel{\textstyle <}{\sim}\;$}}

\begin{document}

\title{Strong decays of higher charmonium states into open-charm meson pairs }
\author{
Long-Cheng Gui$^{1,3}$~\footnote{E-mail: guilongcheng@hunnu.edu.cn}, Long-Sheng Lu$^{1}$, Qi-Fang L\"{u}$^{1,3}$~\footnote {E-mail: lvqifang@hunnu.edu.cn}, Xian-Hui Zhong$^{1,3}$~\footnote {E-mail: zhongxh@hunnu.edu.cn}, and Qiang Zhao$^{2,3,4}$~\footnote{E-mail: zhaoq@ihep.ac.cn}}

%\affiliation{ 2) Institute of High Energy Physics and Theoretical Physics Center for Science Facilities,
%Chinese Academy of Sciences, Beijing 100049, China}

\affiliation{ 1) Department
of Physics, Hunan Normal University, and Key Laboratory of
Low-Dimensional Quantum Structures and Quantum Control of Ministry
of Education, Changsha 410081, China }

\affiliation{ 2) Institute of High Energy Physics and Theoretical Physics Center for Science Facilities,
Chinese Academy of Sciences, Beijing 100049, China}

\affiliation{ 3) Synergetic Innovation Center for Quantum Effects and Applications (SICQEA),
Hunan Normal University, Changsha 410081, China}
\affiliation{ 4)  School of Physical Sciences, University of Chinese Academy of Sciences, Beijing 100049, China}
%\date{\today}

\begin{abstract}
The open-charm strong decays of higher charmonium states up to
the mass of the $6P$ multiplet are systematically studied in the $^3P_0$ model.
The wave functions of the initial charmonium states are calculated in the linear potential (LP)
and screened potential (SP) quark model. The decay widths for most of the
well-established charmonium states above the open-charm thresholds can be reasonably
described. By comparing our quark model calculations with
the experimental observations we also discuss the nature of some of the newly observed charmonium-like
states. It is found that (i) the $\psi(4415)$ may favor the $\psi(4S)$ or
$\psi_1(3D)$ assignment. There may exist two highly overlapping vector
charmonium states around 4.4 GeV; (ii) In the LP model
the $J^{PC}=1^{--}$ $Y(4660)$ resonance and the $J^{PC}=0^{++}$ $X(4500)$ resonance
may be assigned as the $\psi(5S)$ and $\chi_{c0}(4P)$,
respectively; (iii) The newly observed state $X^*(3860)$ can be assigned as
the $\chi_{c0}(2P)$ state with a narrow width of about $30$ MeV;
(iv) It seems to be difficult to accommodate the $X(4140)$ and $X(4274)$ states in the
same potential model as excited $\chi_{c1}$ states.
(v) The $X(3940)$ resonance can be assigned as the $\eta_c(3S)$ state; (vi) The vector charmonium-like
states $Y(4230/4260,4360)$ and scalar $X(4700)$ cannot be described by any conventional
charmonium states self-consistently in our model.

\end{abstract}

%\pacs{12.39.Jh, 13.30.-a, 14.20.Lq}

\maketitle

\section{Introduction}

During the last decade, many new charmonium-like states above/near open-charm
thresholds, such as $X(3940)$, $X(4140/4274)$, $Y(4230/4260)$,
$Y(4360)$, $Y(4660)$, $X(4500)$ and $X(4700)$, have been reported by the
Belle, BaBar, LHCb, BESIII, CLEO Collaborations and so on~\cite{Olive:2016xmw}.
Lately, a new charmonium-like state $X^*(3860)$ was observed in the
$e^+e^-\to J/\psi D\bar{D}$ process by the Belle
Collaboration~\cite{Chilikin:2017evr}. These newly observed charmonium-like states have attracted a lot of attention from the hadron physics community. One obvious feature is that most of them have masses located around $S$-wave open-flavor thresholds and cannot be easily accommodated by
the conventional quark model. Because of this, they have initiated tremendous interests
and different ideas. Detailed reviews on the status of these charmonium-like states can be found in
Refs.~\cite{Chen:2016qju,Olsen:2014qna,Esposito:2016noz,Guo:2017jvc,Olsen:2017bmm,
Lebed:2016hpi,Eichten:2007qx}, where some of these states are categorized as exotic hadrons.

Although there has been progress made during the past years, there still exist many mysteries
to be uncovered. As we have known that exotic states with normal quantum numbers (
e.g. tetraquark states~\cite{Chen:2016qju}, hadronic molecules~\cite{Guo:2017jvc}, charmonium hybrids~\cite{Guo:2008yz}, and so on) can hardly be
distinguished from the normal ones, in order to understand exotic candidates, one should
also have a reasonable description of the normal hadron spectrum. In the charmonium sector the
low-lying states can be very well described by non-relativistic potential quark model such as the
Cornell model~\cite{Eichten:1974af} and Godfrey-Isgur model~\cite{Godfrey:1985xj,Barnes:2005pb,Godfrey:2015dia}. However,
it is realized that the open channel effects would become essential for higher excited states and it
is still challenging to include such effects in a coherent way~\cite{Barnes:2005pb,Li:2009zu}.

In order to understand these recently observed charmonium-like states, a better understanding of
the charmonium spectrum can be regarded as a prerequisite. In this work we take the strategy of
studying systematically the charmonium open flavor decays within the widely used
linear potential (LP) model~\cite{Godfrey:1985xj,Barnes:2005pb,Godfrey:2015dia} and
screened potential (SP) model~\cite{Li:2012vc,Li:2009zu} such that most of the conventional
charmonium states can be identified by comparing with the experimental measurements. Although it
should be recognized that the SP model may not be sufficient for including the full open threshold effects,
we anticipate that unusual phenomena arising from such a study would indicate signals for unconventional
structures of some of those charmonium-like states. In Ref.~\cite{Deng:2016stx}  the charmonium spectrum
and their electromagnetic (EM) transitions have been studied within both LP and SP models.
For the low-lying charmonium states with a mass of $M<4.0$ GeV, both models give comparable predictions.
However, for the higher charmonium states with a mass of $M>4.1$ GeV,
the SP model gives very different results from the LP model.
For example, in the SP model, the $J^{PC}=1^{--}$ charmonium-like states $Y(4260)$ and $Y(4360)$
are good candidates for the $\psi(4S)$ and $\psi_1(3D)$ states, respectively,
while the $J^{PC}=1^{++}$ charmonium-like states $X(4140)$ may be
assigned as the $\chi_{c1}(3P)$ state. In contrast, there is no
room for the charmonium-like states $Y(4260)$, $Y(4360)$ and $X(4140)$
in the LP model, the well-established state $\psi(4415)$ may be
assigned to $\psi(4S)$ or $\psi_1(3D)$, and the charmonium-like states $X(4274)$
seems to be a candidate of $\chi_{c1}(3P)$. Such a result has already shown different dynamic origins introduced by the color screening effects. To clarify the nature of the newly observed charmonium-like states, we continue to investigate the open-flavor strong decays in the SP and LP models in this work. The differences between these two models and their comparisons with experimental observations can provide valuable information on the internal structures of these charmonium-like states.

By adopting the wave functions of the charmonium states calculated with the LP and SP models
in our previous work~\cite{Deng:2016stx} their strong decay amplitudes can be calculated
by the widely used $^3P_0$ model~\cite{Micu:1968mk,LeYaouanc:1972vsx,LeYaouanc:1973ldf,LeYaouanc:1977fsz,
LeYaouanc:1977gm,Arneodo:1979we,Ferretti:2015rsa}.
In this method, one assumes that a $q\bar{q}$ pair is produced from
the vacuum with the vacuum quantum numbers, $J^{PC}=0^{++}$,
and the decay of the charmonium state takes place by regrouping
the new $q\bar{q}$ pair created from the vacuum and the $c\bar{c}$
in the initial state into the outgoing open-charm meson pair via a
rearrangement process.

As an important topic in hadron physics, the open-charm strong decays of
the charmonium states are often discussed in the literature~\cite{LeYaouanc:1977fsz,LeYaouanc:1977gm,
Arneodo:1979we,Eichten:2004uh,Eichten:2005ga,Ding:2007rg,Segovia:2008zz,Yang:2009fj,Wang:2014voa,Wang:2014lml,
Chen:2016iua,Yu:2017bsj,Wang:2016mqb,Jiang:2013epa,Wang:2013lpa,Chen:2014wva,Ortega:2017qmg,Segovia:2012cd,
Gonzalez:2016fsr,Ferretti:2014xqa,Close:1995eu,Barnes:2005pb,Segovia:2013kg,Liu:2009fe,Ferretti:2013faa,Ebert:2014jxa}.
Several pioneering works can be found in
Refs.~\cite{LeYaouanc:1977fsz,LeYaouanc:1977gm,Arneodo:1979we}, where the open charm strong decays of $\psi(3770)$,
$\psi(4040)$, $\psi(4160)$ and $\psi(4415)$ have been evaluated about forty years ago.
Stimulated by first observed charmonium-like resonance $X(3872)$ at Belle~\cite{Choi:2003ue}
and CDF~\cite{Acosta:2003zx}, Eichten, Lane and Quigg analyzed the open charm strong decays of charmonium
states near threshold in the Cornell coupled-channel model~\cite{Eichten:2004uh,Eichten:2005ga}.
In 2007, Ding, Zhu and Yan  considered the open flavor strong decays of $Y(4360)$ and $Y(4660)$ as $3 ^3D_1$ and $5 ^3S_1$
canonical charmonium in the simple harmonic oscillator wave function approximation in the framework of
flux tube model~\cite{Ding:2007rg}. In 2008, Segovia \emph{et al.} calculated the open-flavor strong
decays of the $J^{PC}=1^{--}$ charmonium states in the $^3P_0$ model~\cite{Segovia:2008zz},
where the new $X(4360)$ state was considered to be the $\psi(4S)$ state and the $\psi(4415)$ as
the $\psi_1(3D)$ state, which differs from other assignments. In 2009, as conventional charmonium states,
the open-flavor strong decays of the newly observed resonances $X(3915)$ and $X(4350)$
were studied by Liu \emph{et al.} within the $^3P_0$ model, the strong decay properties
indicate that they may be assigned as $\chi_{c0}(2P)$ and $\chi_{c2}(3P)$,
respectively~\cite{Liu:2009fe}. Further studies of the open-flavor strong decays of
$P$-wave charmonium states were also carried out within the $^3P_0$ model by several groups
in recent years~\cite{Yang:2009fj,Wang:2014voa,Chen:2016iua,Yu:2017bsj}. It is found that
$X(3915)$ may be disfavored the assignment of $\chi_{c0}(2P)$~\cite{Yang:2009fj,Wang:2014voa},
the $X(4140)$ may favor the $\chi_{c1}(3P)$ state~\cite{Chen:2016iua}, while the newly observed state $X^*(3860)$
can be a good candidate of $\chi_{c0}(2P)$ with a broad width~\cite{Yu:2017bsj}.
Recently, the Bethe-Salpeter method was also extended to
deal with the open-charm strong decays of several charmonium states~\cite{Wang:2016mqb,Jiang:2013epa,Wang:2013lpa}.
However, as emphasized earlier, systematic studies of the full spectrum are essential for a better understanding of the underlying dynamics. Furthermore, how to properly treat the strong $S$-wave threshold interactions is a key issue for the description of near-threshold states~\cite{Guo:2017jvc}. The SP model can partially account for such an effect which makes the systematic comparison between the LP and SP model results interesting. Note that the most recent systematic study of the strong decays of higher charmonium
states was carried out by Barnes, Godfrey
and Swanson~\cite{Barnes:2005pb} quite long ago. It is necessary to re-investigate in a systematic way
the strong decays of the higher charmonium states by combining the recent progress in theory
and experiments.

The paper is organized as follows. In Sec.~\ref{3p0},
a brief introduction to the $^3P_0$ strong decay model is presented.
In Sec.~\ref{result}, we focus on the calculation results and discuss the phenomenological consequences in comparison with the experimental data. A summary is given in Sec.~\ref{sum}.

\section{$^3P_0$ Model}\label{3p0}

\begin{figure}[h]
\begin{center}
\centering  \epsfxsize=7.6 cm \epsfbox{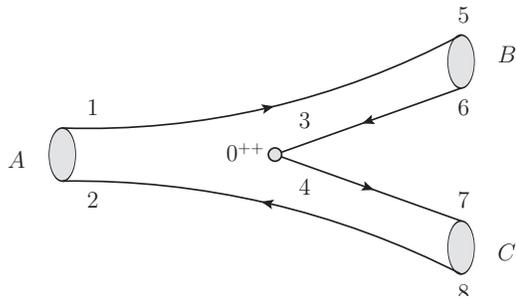} %\epsfxsize=7.6 cm \epsfbox{decay.eps}
\vspace{-0.3 cm} \caption{The strong decay mechanism for meson two-body decays $A\to BC$ in the $^3P_0$ Model.} \label{fig1}
\end{center}
\end{figure}

In this work, we used the $^3P_0$ model to calculate the Okubo-
Zweig-Iizuka (OZI) allowed strong decay widths for the charmonium states above $D\bar{D}$ threshold.
The $^3P_0$ model is a model that describes the quark pair creation mechanism of the OZI allowed strong decays based on the quark model.
It is firstly proposed by Micu~\cite{Micu:1968mk} and then extended by Le Yaouanc {\it et al}~\cite{LeYaouanc:1972vsx,LeYaouanc:1973ldf}.
This model has been widely applied to deal with the open-charm strong decays of the charmonium states~\cite{LeYaouanc:1977fsz,LeYaouanc:1977gm,
Arneodo:1979we,Segovia:2008zz,Yang:2009fj,Wang:2014voa,
Chen:2016iua,Yu:2017bsj,Chen:2014wva,Ortega:2017qmg,Segovia:2012cd,
Gonzalez:2016fsr,Ferretti:2014xqa,Barnes:2005pb,Wang:2014lml}.
In the $^3P_0$ model, one assumes that a quark-antiquark pair is produced from the vacuum with the quantum
number $0^{++}$ and the heavy meson decay takes place via the rearrangement of the four quarks. Such a process is empirically illustrated
in Fig.~\ref{fig1}.

The quark pair creation process from vacuum can be described as,
\begin{eqnarray}
    T & = & -3 \gamma \sqrt{96 \pi} \sum_{m}^{} \langle 1 m 1 -m| 0 0 \rangle \int_{}^{} d\mathbf{p}_3 d\mathbf{p}_4 \delta^3 (\mathbf{p}_3 + \mathbf{p}_4) \nonumber\\
      & \times &  \mathcal{Y}_1^m \left(\frac{\mathbf{p}_3 - \mathbf{p}_4}{2}\right)  \chi_{1-m}^{34}  \phi_0^{34} \omega_0^{34} b_{3i}^\dagger (\mathbf{p}_3) d_{4j}^\dagger (\mathbf{p}_4) \ ,
\end{eqnarray}
where $\gamma$ is a dimensionless constant that denotes the strength of the quark-antiquark pair creation with
momentum $\mathbf{p}_3$ and $\mathbf{p}_4$ from vacuum; $b_{3i}^\dagger (\mathbf{p}_3)$ and $d_{4j}^\dagger(\mathbf{p}_4)$ are the creation operators for the quark and antiquark, respectively; the subscriptions, $i$ and $j$, are the SU(3)-color indices of the created quark and anti-quark;
$\phi_0^{34}=(u\bar u +d\bar d +s \bar s)/\sqrt 3$ and $\omega_{0}^{34}=\frac{1}{\sqrt{3}} \delta_{ij}$ correspond to flavor and
color singlets, respectively; $\chi_{{1,-m}}^{34}$ is a spin triplet
state; and $\mathcal{Y}_{\ell m}(\mathbf{k})\equiv
|\mathbf{k}|^{\ell}Y_{\ell m}(\theta_{\mathbf{k}},\phi_{\mathbf{k}})$ is the
$\ell$-th solid harmonic polynomial. The factor $(-3)$ is introduced for
convenience, which will cancel the color factor.

In the center-of-mass (c.m.) frame of the initial meson $A$, the helicity amplitude can be written as,
\begin{eqnarray}\label{T-matrix}
\mathcal{M}^{M_{J_A}M_{J_B} M_{J_C}}(\mathbf{P})&=&\gamma \sqrt{96 \pi} \sum_{\begin{subarray}{l}
      M_{L_A},M_{S_A},\\
      M_{L_B},M_{S_B},\\
      M_{L_C},M_{S_C},m
      \end{subarray}}
I^{M_{L_A},m}_{M_{L_B},M_{L_C}}({\textbf{P}})\nonumber\\
 && \nonumber   \times\langle
L_A M_{L_A}; S_A M_{S_A} | J_A M_{J_A} \rangle\nonumber\\
&&
\times \langle 1\;m;1\;-m|\;0\;0 \rangle\langle L_B M_{L_B};
S_B M_{S_B} | J_B M_{J_B} \rangle\nonumber\\
&&
\times\langle
L_C M_{L_C}; S_C M_{S_C} | J_C M_{J_C} \rangle\langle\phi^{1
3}_B \phi^{2 4}_C | \phi^{1 2}_A \phi^{3 4}_0
 \rangle\nonumber\\
&&
\times
\langle \chi^{1 3}_{S_B M_{S_B}}\chi^{2 4}_{S_C
M_{S_C}}  | \chi^{1 2}_{S_A M_{S_A}} \chi^{3 4}_{1 -\!m} \rangle
\;,
\end{eqnarray}
with the integral in the momentum space,
\begin{eqnarray}\label{integral}
\begin{aligned}
&I^{M_{L_A},m}_{M_{L_B},M_{L_C}}(\mathbf{P})=\int d^3\mathbf{p}_3 \Psi^*_{n_B L_B M_{L_B}}\left(\frac{m_3 \mathbf{P}}{m_1+m_3}-\mathbf{p}_3 \right) \\
&
\times\Psi^*_{n_C L_C M_{L_C}}\left(\frac{-m_3 \mathbf{P}}{m_2+m_3}+\mathbf{p}_3 \right)\Psi_{n_A L_A M_{L_A}}\left(\mathbf{P}-\mathbf{p}_3 \right)\mathcal{Y}_{1m}(\mathbf{p}_3).
\end{aligned}
\end{eqnarray}
In the above equations, ($J_{A}$, $J_{B}$ and $J_{C}$), ($L_A$, $L_B$ and $L_C$) and ($S_A$, $S_B$ and $S_C$) are the quantum numbers of the total angular momenta, orbital angular momenta and total spin for hadrons $A,B,C$, respectively;
in the c.m. frame of hadron
$A$, the momenta $\mathbf{P}_B$ and $\mathbf{P}_C$ of mesons $B$ and $C$ satisfy
$\mathbf{P}_B=-\mathbf{P}_C\equiv \mathbf{P}$; $m_1$ and $m_2$ are the
constituent quark masses of the initial hadron $A$; $m_3$ is
the mass of the anti-quark created from vacuum;
$\Psi_{n_A L_A M_{L_A}}$, $\Psi_{n_B L_B M_{L_B}}$ and $\Psi_{n_C L_C M_{L_C}}$
are the radial wave functions of hadrons $A$, $B$ and $C$, respectively, in
the momentum space, while $\phi^{12}_A$, $\phi^{13}_B$ and $\phi^{24}_C$ ($\chi^{1 2}_{S_A M_{S_A}}$, $\chi^{1 3}_{S_B M_{S_B}}$
and $\chi^{2 4}_{S_C M_{S_C}}$) are the flavor (spin) wave functions
of hadrons $A$, $B$ and $C$, respectively;
$\langle \phi^{1
3}_B \phi^{2 4}_C | \phi^{1 2}_A \phi^{3 4}_{00}\rangle$ and $\langle  \chi^{1 3}_{S_B M_{S_B}} \chi^{2 4}_{S_C M_{S_C}}|
\chi^{1 2}_{S_A M_{S_A}} \chi^{3 4}_{1 -\!m} \rangle$ are the flavor and spin matrix elements, respectively;
$\langle L_A M_{L_A}; S_A M_{S_A} | J_A M_{J_A} \rangle$ and $\langle
L_B M_{L_B}; S_B M_{S_B} | J_B M_{J_B} \rangle$, $\langle
L_C M_{L_C}; S_C M_{S_C} | J_C M_{J_C} \rangle$ and
$\langle 1\;m;1\;-m|\;0\;0 \rangle$ are the corresponding Clebsch-Gordan
coefficients.

A partial wave amplitude can be obtained by using the Jacob-Wick formula~\cite{Jacob:1959at}
\begin{equation}\label{eq4}
\begin{aligned}
&
{\mathcal{M}}^{J L}(A\rightarrow BC) = \frac{\sqrt{4\pi (2 L+1)}}{2 J_A+1} \!\! \sum_{M_{J_B},M_{J_C}} \langle L 0 J M_{J_A}|J_A M_{J_A}\rangle \\
& \hspace{1cm}
\times  \langle J_B M_{J_B} J_C M_{J_C} | J M_{J_A} \rangle \mathcal{M}^{M_{J_A} M_{J_B} M_{J_C}}({\textbf{K}}),
\end{aligned}
\end{equation}
where $M_{J_A}=M_{J_B}+M_{J_C}$ ,\;$\mathbf{J}\equiv \mathbf{J}_B+\mathbf{J}_C$ and
$\mathbf{J}_{A} \equiv \mathbf{J}_{B}+\mathbf{J}_C+\mathbf{L}$.

Then the strong decay width for a given decay mode of
meson $A$ is given by
\begin{eqnarray}
\Gamma = 2\pi |\textbf{P}| \frac{E_BE_C}{M_A}\sum_{JL}\Big
|\mathcal{M}^{J L}\Big|^2,\label{de}
\end{eqnarray}
where $M_A$ is the mass of the initial hadron $A$, while $E_B$ and $E_C$ stand for the energies of
final hadrons $B$ and $C$, respectively.

When calculating a decay width of a charmonium state, we adopt the numerical
wave function for a charmonium state calculated by the LP and SP models from our previous work~\cite{Deng:2016stx}.
For the emitted charmed mesons in a decay process, such as $D$ and $D^*$, we use
simple harmonic oscillator (SHO) wave functions as an approximation,
\begin{equation}
\label{}
\begin{aligned}
\Psi_{n L M_L} (\mathbf{P})& = \frac{(-1)^n (-i)^L}{\beta^{(3/2)}} \sqrt{\frac{2 n!}{\Gamma(n+L+3/2)}} \left(\frac{P}{\beta}\right)^L  \\
                  & \times e^{-\frac{P^2}{2 \beta^2}} L_n^{L+1/2} \left(\frac{P^2}{\beta^2}\right) Y_{L M_L} (\Omega_P),
\end{aligned}
\end{equation}
where $\beta$ is the universal harmonic oscillator parameter, and
$L_n^{L+1/2} \left(\frac{p^2}{\beta^2}\right)$ is an associated Laguerre
polynomial.

To partly remedy the inadequacy of the nonrelativistic
wave function as the momentum $\mathbf{P}$ increases, a
commonly used Lorentz boost factor $\gamma_f$ is introduced into the
decay amplitudes~\cite{Li:1995si,Zhao:1998fn,Zhong:2007fx,Zhong:2008kd}
\begin{eqnarray}
\mathcal{M}(\mathbf{P})\to \gamma_f \mathcal{M}(\gamma_f \mathbf{P}),
\end{eqnarray}
where $\gamma_f\equiv M_B/E_B$. In most decays, the three momenta $\mathbf{P}$
carried by the final state mesons are relatively small, which
means the nonrelativistic prescription is reasonable and
corrections from the Lorentz boost are not drastic.

In our calculations, we set $m_u=m_d=330$ MeV, $m_s = 450$ MeV and $m_c = 1483$
MeV for the constituent quark masses. The masses
of the well-established hadrons in the final states used in the calculations are
adopted from the PDG~\cite{Olive:2016xmw}.
In the present work, both $\beta$ and the pair creation
strength $\gamma$ are considered as free parameters, which are determined by
fitting the decay widths of the well-established charmonium states
$\psi(3770)$, $\psi(4040)$, $\psi(4160)$ and $\chi_{c2}(2P)$.
If we adopt the wave function of a charmonium state
calculated using the LP model, we obtain $\beta=0.380$ GeV and $\gamma=0.234$.
And if we adopt the wave function of a charmonium state
calculated using the SP model, we have $\beta=0.356$ GeV and $\gamma=0.217$, which are consistent with the LP results.
With these parameters, the decay widths of $\psi(3770)$, $\psi(4040)$, $\psi(4160)$
and $\chi_{c2}(2P)$ can be reasonably described in both LP and SP models (see Table~\ref{Tab01}). The strong decay properties
for the higher charmonium states up to the mass of the $6P$ multiplet have been listed
in Tables~\ref{tabs}-\ref{tabdw2}.

\section{RESULTS AND DISCUSSION}\label{result}

\begin{table}
\caption{\label{Tab01} Open-charm strong decay widths (in MeV) and decay amplitudes (in GeV$^{-1/2}$) for the four established charmonium states $\psi(3770)$, $\psi(4040)$, $\psi(4160)$ and $\chi_{c2}(3927)$. The experimental values are taken from the PDG~\cite{Olive:2016xmw}.}
\begin{tabular}{c   | c  c  c  c  c  c  c  c  c}\hline\hline
\multirow{2}{*}{State} & \multirow{2}{*}{Mode}& \multicolumn{2}{c}{\underline{~LP model~}} &\multicolumn{2}{c}{\underline{~SP model~}} &Amps. (LP/SP)\\
&    &      $\Gamma_{th}$   & Br (\%) & $\Gamma_{th}$    & Br (\%)  &[$\Gamma_{exp}$]    \\ \hline
$\psi(3770)$ & $D^0\bar{D}^0$         & $15$            &$56 $  & $15$             &$56$     &$^1P_1$=0.0921/0.0917 &         \\
               &   $D^+D^-$           & $12$            &$44$   & $12$             &$44$     &$^1P_1$=0.086/0.0863&          \\
               &   Total              & $27$            &100    & $27$             & 100     &[$27\pm1$]           \\ \hline
$\psi(4040)$   &   $D^0\bar{D}^0$     & $1.2$           &$2.0$  & $2.5$            &$ 4.4$   &$^1P_1$=0.0156/0.0225 &           \\
               &   $D^+D^-$           & $0.9$           &$1.6$  & $2.2$            &$ 3.8$   &$^1P_1$=0.0139/0.0211 &          \\
               &   $D^{*0}D^0$        & $4.7$           &$7.9$  & $1.3$            &$ 2.3$   &$^3P_1$=-0.0356/-0.0186&           \\
               &   $D^{*+}D^{-}     $ & $6.0$           &$10 $  & $2.1$            &$38$     &$^3P_1$=-0.041/-0.0244 &         \\
               &   $D^{*0}D^{*0}$     & $23$            &$39 $  & $25$             &$44 $    &$^1P_1$=-0.0277/-0.0285&          \\
               &                      &                 &       &                  &         &$^5P_1$=0.1236/0.1275 &          \\
               &   $D^{*+}D^{*-}$     & $18$            &$30 $  & $19$             &$34 $    &$^1P_1$=-0.0257/-0.0268 &         \\
               &                      &                 &       &                  &         &$^5P_1$=0.1151/0.1199 &          \\
               &   $D_sD_s$           & $5.9$           &$9.0$  & $3.3$            &$ 7.3$   &$^1P_1$=0.0426/0.0377 &          \\
               &   Total              & $60$            & 100   & $55$             &100      &[$80\pm10$]\\ \hline
$\psi(4160)$ & $D^0\bar{D}^0$         & $6.1$           &$7.9$  & $7.0$            &$ 8.4$   &$^1P_1$=0.0312/0.0333&          \\
               &   $D^+D^-$           & $5.9$           &$7.5$  & $6.9$            &$ 8.2$   &$^1P_1$=0.0307/0.0332  &           \\
               &   $D^{*0}D^0$        & $1.4$           &$1.9$  & $3.5$            &$ 4.1$   &$^3P_1$=-0.0166/-0.0257  &           \\
               &   $D^{*+}D^{-}$      & $1.2$           &$1.5$  & $3.1$            &$ 3.8$   &$^3P_1$=-0.015/-0.0246&           \\
               &   $D^{0*}D^{0*}$     & $27 $           &$35 $  & $27$             &$34$     &$^1P_1$=-0.0139/-0.0061&         \\
               &                      &                 &       &                  &         &$^5P_1$=0.0062/0.0027 &          \\
               &                      &                 &       &                  &         &$^5F_1$=0.0818/0.0848 &          \\
               &   $D^{*+}D^{*-}$     & $26 $           &$34$   & $28$             &$33$     &$^1P_1$=-0.0154/-0.0076&        \\
               &                      &                 &       &                  &         &$^5P_1$=0.0069/0.0034 &          \\
               &                      &                 &       &                  &         &$^5F_1$=0.0803/0.0839 &          \\
               &   $D_sD_s$           & $0.6$           &$0.2$  & $0.9$            &$ 0.3$   &$^1P_1$=0.0056/-0.007 &           \\
               &   ${D_s}^*D_s$       & $10$            &$13$   & $2.7$            &$ 7.8$   &$^3P_1$=-0.0548/-0.0454 &           \\
               &   Total              & $79$            & 100   & $80$             &  100    &[$70\pm10$]\\ \hline
$\chi_{c2}(3927)$& $D^0\bar{D}^0$     & $12 $           &$33$   & $12 $          &$30$     &$^1D_2$=-0.057/-0.0566&        \\
               &   $D^+D^-$           & $12 $           &$32$   & $12 $          &$30$     &$^1D_2$=-0.0573/-0.0573 &         \\
               &   $D^{*0}D^0$        & $7.8$           &$21$   & $9.5$           &$23$     &$^3D_2$=-0.0623/-0.0688&         \\
               &   $D^{*+}D^{-}$      & $5.8$           &$15$   & $7.1$            &$17$     &$^3D_2$=-0.0556/-0.0617 &         \\
               &   Total              & $38$            &100    & $41$             &100      &[$24\pm6$] \\ \hline \hline

\end{tabular}
\end{table}

\subsection{Well-established $c\bar{c}$ states}

We choose four states above the $D\bar{D}$ threshold to determine the parameters in our model, i.e. $\psi(3770)$, $\psi(4040)$, $\psi(4160)$,
and $\chi_{c2}(3927)$ which are broadly accepted as
$1^{3}D_{1}$, $3^{3}S_{1}$, $2^{3}D_{1}$, and $2^3P_2$ states, respectively. The first three states, $\psi(3770)$, $\psi(4040)$, and $\psi(4160)$, have been well-established for a long time, while $\chi_{c2}(3927)$ was observed
in experiment quite recently. A good understanding of their
strong decay properties is the starting point for our study of the strong
decay properties of other charmonium states.

\subsubsection{$\psi(3770)$}

The $\psi(3770)$ is assigned to be the $1^{3}D_{1}$ charmonium state though a small $S$-wave component is allowed. This is the first $D$-wave vector charmonium state in the spectrum and located close to the $D\bar{D}$ threshold. In principle, the production of a $D$-wave state will be highly suppressed in $e^+e^-$ annihilations due to the heavy quark spin symmetry (HQSS) constraint. However, as found by experiment, the production cross section for $\psi(3770)$ is actually sizeable. It indicates quite large HQSS breakings in the charmonium sector mainly because the charm quark mass is not heavy enough. As the consequence, its non-$D\bar{D}$ branching ratio turns out to be much larger than a naive estimate based on the HQSS (see e.g. Refs.~\cite{Zhang:2009kr,Zhang:2009gy,Chen:2012qq} for a modern view of this topical issue).

The dominant decay mode of $\psi(3770)$ into $D\bar{D}$ is driven by the $D$-wave component in its wave function. By treating it as
a pure $1^{3}D_{1}$ state as the leading approximation, its strong decay properties can be
well understood within both LP and SP models. Note that the differences of the momentum transfers between the charged and neutral $D\bar{D}$ meson pairs will introduce isospin breaking effects to the $\psi(3770)\to D\bar{D}$ couplings in the $^3P_0$ model. Taking into account such isospin breaking effects, we obtain the partial width
ratio between the two modes $D^0\bar{D}^0$ and
$D^+D^-$,
\begin{equation}
\frac{\Gamma(D^0\bar{D}^0)}{\Gamma(D^+D^-)}\simeq 1.25,
\end{equation}
which is consistent with the world average value $1.26\pm 0.021$ from the PDG~\cite{Olive:2016xmw}.

\subsubsection{$\psi(4040)$}

The $\psi(4040)$ resonance is assigned to be the $3^{3}S_{1}$ charmonium state
in the potential quark model. Four open-charm decay modes $D\bar{D}$, $D\bar{D}^*+c.c.$, $D^*\bar{D}^*$
and $D_s\bar{D}_s$ have been seen in experiment~\cite{Olive:2016xmw}. For convenience, we apply as follows the abbreviations $DD$, $DD^*$, $D^*D^*$ and $D_sD_s$ etc for the corresponding charmed and anti-charmed meson pairs in the final state.
Its OZI allowed two-body strong decays in the LP and SP models are calculated and listed in Table~\ref{Tab01}.

We find that the total width $\Gamma\sim 60$ MeV obtained in this
work is slightly smaller than the world average value $80\pm 10$ MeV.
The $\psi(4040)$ mainly decays into the  $D^*D^*$ channel.
Within the LP model, the partial width ratio
\begin{equation}
\frac{\Gamma(DD)}{\Gamma(DD^*)}\simeq 0.20,
\end{equation}
is consistent with the measured value $0.24\pm 0.05\pm 0.12$ from
the BaBar Collaboration~\cite{Aubert:2009aq}. However, our calculation of
\begin{equation}
\frac{\Gamma(D^*D^*)}{\Gamma(DD^*)}\simeq 3.8
\end{equation}
seems to be much larger than the measured value $0.18\pm 0.14\pm 0.03$ from
the BaBar Collaboration~\cite{Aubert:2009aq}.

In the SP model the partial width ratios are
\begin{equation}
\Gamma(DD):\Gamma(DD^*):\Gamma(D^*D^*)\simeq 1:0.7:10,
\end{equation}
which is very different from the results of the LP model and the calculations of other works
~\cite{Barnes:2005pb,Eichten:2005ga,Segovia:2013kg}.

It should be noted that the decay channel of $\psi(4040)\to D^*D^*$ is quite sensitive to the kinematics due to the limited phase space. Furthermore, the partial width ratios extracted in various models seem not to agree with the data. Interestingly, the dominance of the $D^*D^*$ decay channel is supported by the data for $e^+e^-\to D^*D^*$ from Belle~\cite{Abe:2006fj}, and the recent analyses of Ref.~\cite{Du:2016qcr} by solving the Lippmann-Schwinger equation and lineshape studies in Ref.~\cite{Xue:2017xpu}. It shows that an improved measurement of the exclusive decays of $\psi(4040)$ and reliable extraction of its resonance parameters are needed.

\subsubsection{$\psi(4160)$}

The $\psi(4160)$ is assigned to be the $2^{3}D_{1}$ charmonium state
in the quark model. Four open-charm decay modes $DD$, $DD^*$, $D^*D^*$
and $D_sD_s^*$ have been seen in experiments~\cite{Olive:2016xmw}.
Its OZI allowed two-body strong decays using the wave
functions calculated by the LP and SP models are evaluated, respectively, and the results are listed in Table~\ref{Tab01}.

It shows that the measured width of  $\psi(4160)$,
$\Gamma\simeq 70\pm 10$ MeV, can be well described by both the LP and SP models.
The decay rate of $\psi(4160)$ into $D_sD_s$ is tiny,
which can explain why the $D_sD_s$ mode is not seen in experiment.
In the main decay channels of $\psi(4160)$, our calculation gives
\begin{equation}
\psi(4160)\to D^*D^*>DD >DD^*,
\end{equation}
which is consistent with the result of Ref.~\cite{Barnes:2005pb}.
Using the wave function calculated by the LP model, we find that
\begin{equation}
\Gamma(DD):\Gamma(DD^*):\Gamma(D^*D^*)\simeq 4:1:20 ,
\end{equation}
which is similar to the ratios, $4:2:20$, calculated by
the SP model. However, these two ratios
are very different from the measured
ratios $\Gamma(DD):\Gamma(DD^*):\Gamma(D^*D^*)\simeq 1:17:50$ from
the BaBar Collaboration~\cite{Aubert:2009aq}. Notice that in these $P$-wave decay channels
there exist obvious interfering effects between $\psi(4040)$ and $\psi(4160)$. A coherent
partial wave analysis combining all these exclusive channels seems to be necessary for
extracting the resonance parameters for these two states.
It should be mentioned that the measured ratios cannot also be well understood in some
existing models~\cite{Eichten:2005ga,Segovia:2013kg}.

\subsubsection{$\chi_{c2}(2P)$}

The $X(3927)$ was observed in the $\gamma\gamma\to D\bar{D}$ process
by the Belle~\cite{Uehara:2005qd} and BaBar~\cite{Aubert:2010ab} collaborations, and has been a good candidate for $\chi_{c2}(2P)$. We study its strong decay properties in
both LP and SP models and the results are listed in Table~\ref{Tab01}.

It shows that both models give similar strong decay properties for this state. The total width of $\chi_{c2}(2P)$ is predicted
to be $\Gamma\simeq 40$ MeV, which is close to the upper limit of the measurements.
The $\chi_{c2}(2P)$ dominantly decays into the $DD$ channel, while the decay
rate into the $D^*D$ channel is also sizeable. The branching fraction
$Br[\chi_{c2}(2P)\to D^*D]$ can reach up to $\sim 40\%$.
In the LP model, the partial width ratio is found to be
\begin{equation}
\frac{\Gamma(D^*D)}{\Gamma(DD)}\simeq 0.55,
\end{equation}
which is slightly smaller than the ratio 0.68 obtained in the SP model.
The $D^*D$ decay mode can be searched in $e^+e^-\to \gamma D^*D$ which is also the channel accessible for $X(3872)$ as predicted in Ref.~\cite{Guo:2013nza}.

\begin{table*}
\caption{\label{tabs} Open-charm strong decay properties for the $S$-wave charmonium states in the LP and SP models.
The widths listed in the brackets are calculated with the masses of observations.}
\begin{tabular}{c | c  c  c  c  c  c  c  l  c  c}\hline\hline
\multirow{2}{*}{State} & Decay& \multicolumn{4}{c}{\underline{~~~~~~~~~~~~~~~~~~~~~~~~~~~~LP model~~~~~~~~~~~~~~~~~~~~~~~~~~~~}}      &\multicolumn{2}{c}{\underline{~~~~~SP model~~~~~}} &Amps. (GeV$^{-1/2}$)& Mass (MeV) \\
 & mode & $\Gamma_{th}$ (MeV) &  Br (\%)& $\Gamma_{th}$ (MeV)~\cite{Barnes:2005pb}&Br(\%)~\cite{Barnes:2005pb}   &$\Gamma_{th}$ (MeV) &  Br (\%)& ~~~~~~LP/SP & LP/SP \\
 \hline
$\eta_{c}(3S)$ & $D^{*}D$        & $21  $~~~[79]       &$28$     &$47 $ & 59   & $5.7$~~~[40]      &$100$    &$^3P_0$=0.0749/ 0.0255&$4048/4004$  & \\
               & $D^{*}D^{*}$    & $54  $~~~[...]      &$72$     &$33 $ & 41   & ...  ~~~...       &...      &$^3P_0$=0.1838/ 0.0339&  $[3940/3940]\footnote{Observed mass from the PDG~\cite{Olive:2016xmw}.}$ &   \\
               & Total           & $75  $~~~[79]      &100      &$80 $ &100   & $5.7$~~~[40]  &100     &        \\ \hline
$\psi(4S)$     & $DD$            & $1.7 $             &$2.6$     &$0.4$ &  0.5 & $2.3$         &$ 19$   &$^1P_1$=0.0145/ 0.0181&$4415/4281$ \\
               & $D^{*}D$        & $1.0 $             &$1.5$     &$2.3$ &  2.9 & $1.0$         &$  7.8$ &   $^3P_1$=0.008/ 0.0126     \\
               & $D^{*}D^{*}$    & $3.8 $             &$5.7 $    &$16 $ & 21   & $5.8$         &$  47$  &   $^1P_1$=-0.0054/ -0.0074   \\
               &                 &                    &         &      &      &               &        &   $^5P_1$=0.0239/ 0.0332      \\
               & $D_sD_s$        & $0.09$             &$ 0.1$   &$1.3$ &  1.6 & $0.1$         &$   0.8$&   $^1P_1$=0.0035/ 0.0042     \\
               & $D_s^*D_s$      & $3.3 $             &$ 5.0$   &$2.6$ &  3.3 & $3.1 $        &$  26$  &   $^3P_1$=0.0239/ 0.0269       \\
               & $D_s^*D_s^*$    & $3.8 $             &$ 5.7$   &$0.7$ &  0.8 & $0.05$        &$   0.4$&   $^1P_1$=0.0064/ -0.001    \\
               &                 &                    &         &      &      &               &        &   $^5P_1$=-0.0286/ 0.0047    \\
               & $DD_{1}$        & $12 $              &$18 $    &$31 $ & 40   & $...$         & ...    &   $^3S_1$=0.00004/ ... \\
               &                 &                    &         &      &      &               &        &   $^3D_1$=0.0578/ ... \\
               & $DD_{1}^{'}$    & $16 $              &$24$     &$1.0$ &  1.2 & $...$         & ...    &   $^3S_1$=-0.0678/ ...\\
               &                 &                    &         &      &      &               &        &   $^3D_1$=0.00004/ ... \\
               & $DD_{2}$        & $17 $              &$25$     &$23 $ & 29   & $...$         & ...    &   $^5D_1$=-0.0746/ ...\\
               & $D^{*}D_{0}$    & $8.7 $             &$13$     &$0.0$ &  0.0 & $...$         & ...    &   $^3S_1$=0.0531/ ... \\
               & Total           & $67  $             & 100     &$78 $ &100   &$12 $          &  100   &   & \\ \hline
$\eta_{c}(4S)$ & $D^{*}D$        & $0.3 $             &$   0.5$ &$6.3$ &10.3  &$0.5 $         &$ 2.8$  &   $^3P_0$=0.0067/ -0.0095&$4388/4264$  \\
               & $D^{*}D^{*}$    & $5.6 $             &$   8.5$ &$14 $ &22.9  &$6.5$          &$  34$  &   $^3P_0$=0.0303/ 0.0368   \\
               & ${D_s}^*D_s$    & $5.7 $             &$   8.7$ &$2.2$ &3.6   &$4.6$          &$  24$  &   $^3P_0$=-0.0321/ -0.0332 \\
               & $D_s^*D_s^*$    & $1.6 $             &$   2.5$ &$2.2$ &3.6   &$0.4$          &$ 1.9$  &   $^3P_0$=-0.0199/ 0.0136 \\
               & $DD_{0}$        & $24 $              &$  36.7$ &$11 $ &18.0  &$7.4$          &$  38$  &   $^1S_0$=0.0731/ 0.0521  \\
               & $DD_{2}$        & $28 $              &$  43.1$ &$24 $ &39.3  &$...$          & ...    &   $^5D_0$=-0.1068/ ...   \\
               &      Total      & $66 $              &   100   &$61 $ &100   &$19$           & 100    &    \\ \hline
$\psi(5S)$     & $DD$            & $0.6~$    ~~[0.3]  &$   1.1$ & ...  & ... &$1.1$~~[0.6]   &$   6.6$&    $^1P_1$=0.0075/ 0.0113  & $4711/4472$     \\
               & $D^{*}D$        & $1.2~$    ~~[0.2]  &$   2.1$ & ...  & ... &$1.5$~~[0.3]   &$   8.8$&    $^3P_1$=0.0112/ 0.0137  & $[4643/4415]^a$ \\
               & $D^{*}D^{*}$    & $0.0~$~~[0.7]      &    0.0  & ...  & ... &$0.2$ ~~[1.4]  &$   0.9$&    $^1P_1$=0.0001/ -0.001   \\
               &                 &                    &         &      &     &               &        &    $^5P_1$=0.0003/ 0.0045    \\
               & $D_sD_s$        & $0.0~$~~[0.04]     &    0.0  & ...  & ... &$0.0$ ~~[0.04] & 0.0    &    $^1P_1$=0.0002/ -0.0001   \\
               & ${D_s}^*D_s$    & $0.3~$   ~~[0.6]   &$   0.5$ & ...  & ... &$0.5 $~~[0.9]  &$   3.3$&    $^3P_1$=0.055/0.0092     \\
               & ${D_s}^*D_s^*$  & $1.5~$   ~~[1.4]   &$   2.5$ & ...  & ... &$1.3$ ~~[0.5]  &$   7.8$&    $^1P_1$=0.003/0.0035     \\
               &                 &                    &         &      &     &               &        &    $^5P_1$=-0.0134/-0.0155   \\
               & $DD_{1}$        & $0.06$   ~~[0.6]   &$   0.1$ & ...  & ... &$1.0 $~~[1.6]  &$   6.1$&    $^3S_1$=0.00002/0.00002       \\
               &                 &                    &         &      &     &               &        &    $^3D_1$=-0.003/0.0152    \\
               & $DD_{1}^{'}$    & $10~~$    ~~[10 ]  &$  17.5$ & ...  & ... &$5.2$ ~~[1.7 ] &$  31$  &    $^3S_1$=-0.038/-0.0349   \\
               & $DD_{2}$        & $0.12$   ~~[1.6]   &$   0.2$ & ...  & ... &$1.7$ ~~[1.0]  &$   9.9$&    $^5D_1$=-0.0041/-0.0206    \\
               & $D^{*}D_{0}$    & $11~~$   ~~[10 ]   &$  18.3$ & ...  & ... &$3.8$ ~~[0.5 ] &$  23$  &    $^3S_1$=0.0392/0.0308     \\
               & $D^{*}D_{1}$    & $4.3~$   ~~[5.7]   &$   7.3$ & ...  & ... &$0.2 $~~[...]  &$   1.4$&    $^3S_1$=-0.00003/0.0       \\
               &                 &                    &         &      &     &               &        &    $^3D_1$=0.0134/0.0052     \\
               &                 &                    &         &      &     &               &        &    $^5D_1$=0.0232/0.009     \\
               & $D^{*}D_{1}^{'}$& $18~~$   ~~[8.3]   &$  30.3$ & ...  & ... &$0.3$ ~~[...]  &$   1.5$&    $^3S_1$=-0.0551/-0.0113   \\
               &                 &                    &         &      &     &               &        &    $^5D_1$=-0.00002/0.0  \\
               & $D^{*}D_{2}$    & $12~~$   ~~[7.8]   &$  20.1$ & ...  & ... &$0.0$ ~~[...]  & 0.0    &    $^3D_1$=0.0073/-0.0001    \\
               &                 &                    &         &      &     &               &        &    $^5D_1$=-0.094/0.0002    \\
               &                 &                    &         &      &     &               &        &    $^7D_1$=-0.0445/0.0009    \\
               & Total           & $58~~$   ~~[47 ]   &   100   & ...  & ... &$17$  ~~[8.4]  &  100   &           \\ \hline
$\eta_{c}(5S)$ & $D^{*}D$        & $0.5 $             &$   0.7$ & ...  & ... &$1.7$          &$ 11$   &$^3P_0$=-0.0075/-0.0147& $4690/4459$  \\
               & $D^{*}D^{*}$    & $0.4 $             &$   0.6$ & ...  & ... &$0.3$          &$ 2.0$  &    $^3P_0$=0.0067/0.0065     \\
               & ${D_s}^*D_s$    & $0.8 $             &$   1.3$ & ...  & ... &$0.9 $         &$ 6.4$  &    $^3P_0$=-0.0099/-0.0123  \\
               & ${D_s}^*{D_s}^*$& $1.4$              &$   2.1$ & ...  & ... &$1.0$          &$ 6.6$  &    $^3P_0$=-0.0134/-0.014  \\
               & $DD_{0}$        & $13 $              &$  18.9$ & ...  & ... &$7.1$          &$  48$  &    $^1S_0$=0.0404/0.0365    \\
               & $DD_{2}$        & $2.0 $             &$   3.0$ & ...  & ... &$3.4$          &$  23$  &    $^5D_0$=-0.0176/-0.0301  \\
               & $D_1D^*$        & $11  $             &$  15.9$ & ...  & ... &$0.07$         &$ 0.5$  &    $^1S_0$=-0.00004/0.0      \\
               &                 &                    &         &      &     &               &        &    $^5D_0$=-0.0433/-0.0062  \\
               & $D^*D'_1$       & $26  $             &$  41.3$ & ...  & ... &$0.4 $         &$ 2.9$  &    $^1S_0$=0.0706/-0.0168    \\
               &                 &                    &         &      &     &               &        &    $^1S_0$=0.00003/0.0    \\
               & $D^{*}D_{2}$    & $11  $             &$  16.1$ & ...  & ... &...            & ...    &    $^5D_0$=-0.0455/    \\
               &Total            & $67$               &  100    & ...  & ... &$15$           & 100    &    \\ \hline\hline
\end{tabular}
\end{table*}

\subsection{Candidates of higher $c\bar{c}$ states with $J^{PC}=1^{--}$}

The higher vector charmonium states,
$\psi(4S)$, $\psi(5S)$ and $\psi_1(3D)$, are still not well-established.
During the past decade, several $J^{PC}=1^{--}$ charmonium-like states,
$Y(4230,4260)$, $Y(4360)$ and $Y(4660)$, have been observed
in experiment~\cite{Olive:2016xmw}. Some of them exhibit unusual properties that are very different
from the expectations as conventional $c\bar{c}$ states. In addition, although $\psi(4415)$ has been
well-established in experiment, its structure and quark model assignment still need to be studied.

\subsubsection{ $Y(4230,4260)$}\label{Y-4260}

The $Y(4260)$ state turns out to be a mysterious state from the very beginning. It was first reported by
the BaBar Collaboration in the initial state radiation $e^+e^-\to \gamma_\text{ISR} J/\psi\pi^+\pi^-$~\cite{Aubert:2005rm},
and then confirmed by CLEO-c~\cite{He:2006kg} and Belle~\cite{Yuan:2007sj} experiments in the same channel.
However, its presence in open charm decay channels is not obvious at all, which has provoked a
lot of theoretical interpretations in the literature. Comprehensive reviews can be found in several
recent review articles~\cite{Chen:2016qju,Olsen:2014qna,Guo:2017jvc,Olsen:2017bmm,Lebed:2016hpi}.
Recently, following the discovery of charged charmonium-like state $Z_c(3900)$ in $e^+e^-\to J/\psi\pi\pi$
at the c.m. energy of 4.26 GeV~\cite{Ablikim:2013mio}, the BESIII Collaboration observed more detailed
structures around the $Y(4260)$ in several exclusive decay channels, namely, $J/\psi\pi\pi$~\cite{Ablikim:2016qzw},
$h_c\pi\pi$~\cite{BESIII:2016adj}, $\omega\chi_{c0}$~\cite{Ablikim:2014qwy,Ablikim:2015uix},
and $D^0D^{*-}\pi^++c.c.$~\cite{Gao:2017sqa}. In particular, in Ref.~\cite{Gao:2017sqa} by treating
these two structures around 4.23 and 4.29 GeV as from two Breit-Wigner states, a narrow resonance $Y(4230)$
and a relatively broad resonance $Y(4260)$ are extracted.

However, as studied in a series of works in Refs.~\cite{Wang:2013cya,Cleven:2013mka,Wang:2013kra,Wu:2013onz,Qin:2016spb,Xue:2017xpu},
the energy region of $Y(4260)$ is close to the first narrow $S$-wave open charm threshold $DD_1(2420)$. Therefore,
a strong near-threshold $S$-wave coupling can dynamically generate molecular state, which can then mix with
nearby vector charmonium state and cause nontrivial near-threshold structures. In such a hadronic molecule
scenario, the structures observed in these exclusive decay channels can be accounted for by the
dynamics introduced by the $DD_1(2420)$ threshold.

The interesting phenomena arising from the 4.26 energy region is also a challenge
for potential quark models in the study of the $c\bar{c}$ spectrum. In the LP model,
it is almost impossible to accommodate the mass of $Y(4230,4260)$ in the spectrum.
In contrast, in the SP model the mass of $\psi(4S)$ is found to be around 4.28 GeV.
This might indicate the important role played by the $S$-wave threshold of $DD_1(2420)$
which can be partially accounted for by the screening effects. Therefore, if we assume
that $Y(4260)$ is dominated by the $\psi(4S)$ component in the wave function, we can
investigate its decay properties as a charmonium state. As shown by the results listed
in Table~\ref{tabs}, we find that $Y(4260)$ should be a very narrow state with a width of $\sim 14$ MeV,
and its strong decays are dominated by the $D^*D^*$ mode. Although such a result is
consistent with that from Ref.~\cite{He:2014xna} and several
consequent works~\cite{Chen:2017uof,Chen:2015bma,Chen:2014sra} with the assignment of $Y(4260)$ as the $\psi(4S)$
state, the narrow width does not agree with the measured value of about
$55\pm 19$ MeV~\cite{Olive:2016xmw}. Note that such a solution also cannot be accommodated by the two-state fitting performed by Ref.~\cite{Gao:2017sqa}. Another issue is that if we assign $Y(4260)$ to $\psi(4S)$,
we will be unable to understand the decay properties of $\psi(4415)$ in the SP model at all, which will
be discussed later.

\subsubsection{ $\psi(4415)$}\label{psi-4415}

In the LP model the calculated masses
of $\psi(4S)$ and $\psi_1(3D)$ are located around $\sim 4.4$ GeV.
Thus, the $\psi(4415)$ might be a good candidate
for $\psi(4S)$ or $\psi_1(3D)$ as broadly discussed in the literature.
In contrast, in the SP model, the $\psi(4415)$ is suggested
to be a candidate of $\psi(5S)$. We discuss these possibilities below with details.

Assuming $\psi(4415)$ as the $\psi(4S)$ state in the LP model, the calculated decay widths are
listed in Table~\ref{tabs}. It shows that the predicted total width, $\Gamma\simeq 67$ MeV, is consistent
with the measured value $62\pm 20$ MeV~\cite{Olive:2016xmw}.
The main decay modes are $DD_2$, $DD_1'$, $DD_1$ and
$D^*D_0$~\footnote{In this work, $D_1$ and $D_1'$ stand for the narrow state
$D_1(2420)$ and broad state $D_1(2430)$ listed in the PDG~\cite{Olive:2016xmw}, respectively,
which are considered to be mixed states via $^3P_1$-$^1P_1$ mixing as defined in Ref.~\cite{Zhong:2008kd}.
While $D_0$ and $D_2$ stand for the states $D_0(2400)$ and $D_2(2460)$ listed in the PDG~\cite{Olive:2016xmw}, respectively.
}. The branching fractions of $Br[\psi(4415)\to
DD_2, \ DD_1', \ DD_1, \ D^*D_0]$ are $\mathcal{O}(10\%)$.
Our results for some channels are significantly different from those in~\cite{Barnes:2005pb}.
This difference is mainly due to the different wave functions of initial states.
For example, in the $\psi(4415) \to D^* D_0^*$ case,
if the wave function of initial state $\psi(4415)$ in Ref.~\cite{Barnes:2005pb} is replaced by that of our LP model,
we get a larger partial wave amplitude $^3S_1 = -0.1426$ GeV$^{-1/2}$
than $^3S_1 = -0.00087 $ GeV$^{-1/2}$ obtained in Ref.~\cite{Barnes:2005pb}.
This reflects the fact that the integral part $I^{M_{L_A},m}_{M_{L_B},M_{L_C}}(\mathbf{P})$ is sensitive to
the node position of the wave function of initial states.

In particular, the branching fraction of $Br[\psi(4415)\to D D_2]\sim \mathcal{O}(10\%)$ is consistent with the observations of Belle Collaboration~\cite{Pakhlova:2007fq}.
The decay rates into $D^*D^*$, $DD$, $D^*D$ and $D^*_sD_s$
are relatively small with typical branching fractions $\mathcal{O}(1\%)$. Our calculation result for the decay rate of
$\psi(4415)\to D_sD_s$ is tiny, i.e.,
$Br[\psi(4415)\to D_sD_s]< 10^{-4}$. The present data for $e^+e^-\to D_s^{*}D_s^{*}$ are still with large uncertainties~\cite{Pakhlova:2010ek} though some hints of enhancement around 4.415 seem to be present. Further improved measurement is strongly recommended.

In the LP model the partial width ratio between the $DD$ and $D^*D^*$ channel is
\begin{equation}
\frac{\Gamma(DD)}{\Gamma(D^*D^*)}\simeq 0.45,
\end{equation}
which is close to the upper limit $0.29$ measured by BaBar collaboration~\cite{Aubert:2009aq}. The ratio between $D^*D$ and $D^*D^*$,
\begin{equation}
\frac{\Gamma(D^*D)}{\Gamma(D^*D^*)}\simeq 0.26,
\end{equation}
is also consistent with the measured value $0.17\pm 0.28$ within uncertainties.

Assuming $\psi(4415)$ as the $\psi_1(3D)$ state in the LP model, the calculation results are
listed in Table~\ref{tabdw2}. It is found that the calculated total width,
$\Gamma\sim 60$ MeV, is compatible with the measured value of $\psi(4415)$.
The decays of $\psi_1(3D)$ are governed by $DD_1$. The
branching fraction of $Br[\psi(4415)\to DD_1]$ can reach up to 50\%.
The decay rates into $D^*D^*$, $D^*D_0$, $DD_1'$ and $DD_2$ are sizeable and with the branching fractions of $Br[\psi(4415)\to D^*D^*, \ D^*D_0, \ DD_1', \ DD_2]$ at $\mathcal{O}(10\%)$. The partial width ratio
\begin{equation}
\frac{\Gamma(DD)}{\Gamma(D^*D^*)}\simeq 0.18,
\end{equation}
is in the range of $0.14\pm 15$ measured by BaBar Collaboration~\cite{Aubert:2009aq},
while the partial width radio
\begin{equation}
\frac{\Gamma(D^*D)}{\Gamma(D^*D^*)}\simeq 0.03,
\end{equation}
is also in the range of data $0.17\pm 0.28$~\cite{Aubert:2009aq}.

Finally, we consider the possibility of $\psi(4415)$
as the $\psi(5S)$ state in the SP models~\cite{Li:2009zu,Deng:2016stx}. This is based on the assignment that  in the SP model the mass of $\psi(4S)$ is found to be around 4.28 GeV as investigated in Subsection~\ref{Y-4260}.
With this hypothesis, the strong decay properties of $\psi(4415)$ are calculated and listed in Table~\ref{tabs}.
The results from such an assignment turns out to be inconsistent with the
observations of $\psi(4415)$.

In brief, it shows that the present data cannot distinguish the assignments of $\psi(4415)$ as $\psi(4S)$
or $\psi_1(3D)$ in the LP model, while its assignment as the $\psi(5S)$ in the SP model cannot be supported.
It should be mentioned that at the mass of 4.4 GeV, the nearby $S$-wave open-threshold may
introduce coupled-channel effects of which if the interaction
is strong enough, it can dynamically generate poles in a unitarized formulation and mix with the
charmonium state. A recent study of the dynamic effects arising from
the nearby $D_{s0}(2317)D_s^*$ and $D_{s1}(2460)D_s$ thresholds and their impact on the property of $\psi(4415)$
can be found in Ref.~\cite{Cao:2017lui}. Taking into account that both the $\psi(4S)$ and $\psi_1(3D)$ states
are likely located within this energy region, it is also possible that there may exist two highly
overlapping vector charmonium states around 4.4 GeV. More accurate measurements and more observables
are needed in order to understand the vector spectrum above 4 GeV in the future.

\subsubsection{ $Y(4360)$}

In the vector charmonium spectrum, the $Y(4360)$ resonance was first reported
by the BaBar Collaboration in $e^+e^-\to \psi(2S)\pi^+\pi^-$~\cite{Aubert:2007zz}.
Later, the Belle Collaboration confirmed this state in the same channel~\cite{Wang:2007ea}.
The interesting feature about $Y(4360)$ is its presence in the hidden charm decay channel
but seems to be absent from the open charm decays. This is very similar to $Y(4260)$ when it
was first observed in $e^+e^-\to J/\psi\pi\pi$. This mysterious state has also initiated many
theoretical studies with different possible solutions~\cite{Chen:2016qju,Guo:2017jvc,Lebed:2016hpi}
including possible open-charm effects which can mix and shift the nearby charmonium state.

In the SP model, the mass of $\psi_1(3D)$ is estimated to be $~4.32$ GeV.
Considering only the mass position, $Y(4360)$ can be a good candidate of the $\psi_1(3D)$.
Our calculation results are
listed in Table~\ref{tabdw2}. It shows that
the $\psi_1(3D)$ resonance should be a very narrow state
with a width of $\Gamma\simeq 20$ MeV, and its decays should
be dominated by the $D^*D^*$, $DD$ and $DD_1$ modes. In contrast with the experimental value of the width, i.e. $\Gamma=102\pm 9$ MeV, the calculated
width of $\psi_1(3D)$ is too small. Moreover, it has not been observed in open-charm decay channels. Note that the SP model has partly included the open-charm effects, the mismatching of $Y(4360)$ as the $\psi_1(3D)$ state with the experimental measurement has reflected some unusual properties of $Y(4360)$.

\subsubsection{ $Y(4660)$}

The $Y(4660)$ was observed in association with $Y(4360)$ by the Belle Collaboration
in $e^+e^-\to \psi(2S)\pi\pi$~\cite{Wang:2007ea}. Its assignment is still controversial
though by filling the lower states with some of these observed enhancements,
it leaves the $\psi(5S)$ as a possible option. However, it should be pointed out that
the $S$-wave open-charm threshold $D^*D_2$ is located nearby. Therefore, possible contributions
from the open-channel effects or dynamically generated state cannot be ruled out~\cite{Guo:2017jvc}.

In the SP model, the mass of $\psi(5S)$ is lower than 4.6 GeV and in Subsection~\ref{psi-4415}
the assignment of $\psi(4415)$ as the $\psi(5S)$ has been discussed. The results show that
$\psi(4415)$ does not favor such an assignment.
In the LP model, the mass of $\psi(5S)$ is predicted to be 4711 MeV, which is about 50 MeV larger
than the mass of $Y(4660)$. As the nearest state we investigate its
strong decay properties as the $\psi(5S)$, and the results are listed in Table~\ref{tabs}.

It shows that the calculated width $\sim 50$ MeV is close to the measured value
$\Gamma=70\pm 11$ MeV. The main open-charm decay channels include $DD_1'$,
$D^*D_0$, $D^*D_1$, $D^*D_1'$ and $D^*D_2$ with the branching fractions at the order of $10-20\%$. As mentioned earlier, the $Y(4660)$ was observed in $e^+e^-\to \psi(2S)\pi\pi$ instead of open-charm decay channels. Therefore, further experimental studies confirming or denying its contributions to these open-charm decay channels should be essential for determining its nature.

\begin{table*}
\caption{\label{tab3} Open-charm strong decay properties for the $2P$ and $3P$ charmonium states in the LP and SP models.
The widths listed in the brackets are calculated with the masses of observations. }
\begin{tabular}{c | c  c  c  c  c  cclcccccccccc }\hline\hline
\multirow{2}{*}{State} & Decay& \multicolumn{4}{c}{\underline{~~~~~~~~~~~~~~~~~~~~~~~~~~~~LP model~~~~~~~~~~~~~~~~~~~~~~~~~~~~}}      &\multicolumn{2}{c}{\underline{~~~~~SP model~~~~~}} &Amps. (GeV$^{-1/2}$) & Mass (MeV) \\
 & mode & $\Gamma_{th}$ (MeV) &  Br (\%)& $\Gamma_{th}$ (MeV)~\cite{Barnes:2005pb}&Br(\%)~\cite{Barnes:2005pb}   &$\Gamma_{th}$ (MeV) &  Br (\%)& LP/SP & LP/SP  \\
\hline\hline
$\chi_{c1}(2P)$& $D^{0*}\bar{D}^0$  & $48$              &47    &...    &... &$60 $          & 47   &$^3S_1$=0.1365/0.3192&$3937/3914$     \\
               &                    &                   &      &       &    &               &      &$^3D_1$=-0.0531/-0.00002    \\
               & $D^{+*}D^{-*}$      & $54$              &53    &...    &... &$67$           & 53   &$^3S_1$=0.1542/0.1784      \\
               &                    &                   &      &       &    &               &      &$^3D_1$=-0.0453/-0.0432    \\
               &Total               & $102$             &100   &$165$  &100 &$127  $        & 100  &                          & \\ \hline
$\chi_{c0}(2P)$& $D^0\bar{D}^0$     & $9.3$             &42    &...    &... &$12  $         & 43   &$^1S_0$=-0.0541/-0.0635&$3869/3848$ &\\
\hline\hline
\end{tabular}
\end{table*}

\begin{table*}
\caption{\label{amp4} Open-charm strong decay widths and amplitudes for the $4P$ charmonium states in the LP and SP models.
The widths listed in the brackets are calculated with the masses of observations.}
\begin{tabular}{c | c c c l || c | c c c l c c}\hline\hline
\multirow{2}{*}{State}&\multirow{2}{*}{Mode}& \multicolumn{2}{c}{\underline{~~Width (MeV)~~}} &~~~\underline{Amps.  (GeV$^{-1/2}$)}~~~&\multirow{2}{*}{State}&\multirow{2}{*}{Mode}& \multicolumn{2}{c}{\underline{~~Width (MeV)~~}} &~~~\underline{Amps. (GeV$^{-1/2}$)}~~~ \\
&          &LP  &   SP   & ~~~~~~~~~LP/SP~~~~    &    &      &LP       &   SP                            &   ~~~~~~~~~LP/SP~~~~             \\ \hline
$\chi_{C2}(4P)$   &$DD$              &3.23 & 3.34  &$^1D_2$=0.0181/-0.0202 & $\chi_{c1}(4P)$   &$D^*D$            &5.39 & 3.63  &$^3S_1$=0.0142/-0.0074 & \\
$4614\footnote{Mass (MeV) from the LP model~\cite{Deng:2016stx}.}/4416\footnote{Mass (MeV) from the SP model~\cite{Deng:2016stx}.}$                  &$D^*D$            &7.53 & 7.1   &$^3D_2$=0.0288/-0.0144 & $4590^a/4402^b$    &                  &     &       &$^3D_1$=0.0202/-0.039  & \\
                  &$D^*D^*$          &2.39 & 1.52  &$^1D_2$=0.06/-0.0053   &                   &$D^*D^*$          &3.07 & 0.84  &$^5D_1$=-0.0196/0.0116 & \\
                  &                  &     &       &$^5S_2$=0.002/0.0039   &                   &$D_s^*D_s$        &0.74 & 2.17  &$^3S_1$=0.0097/-0.0193 & \\
                  &                  &     &       &$^5D_2$=-0.0158/0.014  &                   &                  &     &       &$^3D_1$=-0.0016/-0.0030& \\
                  &$D_sD_s$          &0.43 & 0.46  &$^1D_2$=-0.007/0.0081  &                   &$D_s^*D_s^*$      &0.79 & 1.21  &$^5D_1$=-0.011/0.0168  & \\
\hline
\end{tabular}
\end{table*}

\begin{table*}
\caption{\label{amp6} Open-charm strong decay widths and amplitudes for the $5P$ charmonium states in the LP and SP models.
The widths listed in the brackets are calculated with the masses of observations.}
\begin{tabular}{c | c c c l || c | c c c l c c}\hline\hline
\multirow{2}{*}{State}&\multirow{2}{*}{Mode}& \multicolumn{2}{c}{\underline{~~Width (MeV)~~}} &~~~\underline{Amps.  (GeV$^{-1/2}$)}~~~&\multirow{2}{*}{State}&\multirow{2}{*}{Mode}& \multicolumn{2}{c}{\underline{~~Width (MeV)~~}} &~~~\underline{Amps. (GeV$^{-1/2}$)}~~~ \\
&          &LP  &   SP   & ~~~~~~~~~LP/SP~~~~    &    &      &LP       &   SP                            &   ~~~~~~~~~LP/SP~~~~             \\ \hline
$\chi_{c2}(5P)$   & $DD$              & 1.06 & 1.71  &$^1D_2$=-0.0094/-0.0134 & $\chi_{c1}(5P)$   & $D^*D$            & 2.05 &3.03  &$^3S_1$=-0.0073/-0.0122 &\\
$4892\footnote{Mass (MeV) from the LP model~\cite{Deng:2016stx}.}/4575\footnote{Mass (MeV) from the SP model~\cite{Deng:2016stx}.}$   & $D^*D$            & 2.99 & 3.96  &$^3D_2$=-0.0161/-0.0213 &  $4870^a/4563^b$  &                   &      &      &$^3D_1$=-0.0113/-0.0154 &\\
                  & $D^*D^*$          & 1.11 & 1.26  &$^1D_2$=-0.0036/-0.0044 &                   & $D^*D^*$          & 1.55 &1.94  &$^5D_1$=0.0121/0.0158   &\\
                  &                   &      &       &$^5S_2$=-0.0018/-0.0021 &                   & $D_s^*D_s$        & 0.21 &0.26  &$^3S_1$=-0.0042/-0.0056 &\\
                  &                   &      &       &$^5D_2$=0.0093/0.0117   &                   &                   &      &      &$^3D_1$=0.0016 /0.0021  &\\
                  & $D_sD_s$          & 0.1  & 0.25  &$^1D_2$=0.003/0.0055    &                   & $D_s^*D_s^*$      & 0.09 &0.21  &$^5D_1$=0.0031/0.0058   &\\
\hline
\end{tabular}
\end{table*}

\begin{table*}
\caption{\label{amp8} Open-charm strong decay widths and amplitudes for the $6P$ charmonium states in the SP models.
The widths listed in the brackets are calculated with the masses of observations.}
\begin{tabular}{c | c c l || c | c c  l c c}\hline\hline
State  & Mode & Width (MeV) & Amps. (GeV$^{-1/2}$) &State& Mode & Width (MeV)  & Amps. (GeV$^{-1/2}$)  \\ \hline
$\chi_{c2}(6P)$   & $DD$              & 0.91   &$^1D_2$=-0.0219   &$\chi_{c1}(6P)$   & $D^*D$            & 1.16   &$^3S_1$=-0.0337   & \\
$4696\footnote{Mass (MeV) from the SP model~\cite{Deng:2016stx}.}$          & $D^*D$            & 2.26   &$^3D_2$=-0.0398   & $4687^a$         &                   &        &$^3D_1$=-0.0322   & \\
                  & $D^*D^*$          & 0.89   &$^1D_2$= -0.012   &                  & $D^*D^*$          & 0.73   &$^5D_1$=0.0447    & \\
                  &                   &        &$^5S_2$=-0.0284   &                  & $D_s^*D_s$        & 0.27   &$^3S_1$=0.0022    & \\
                  &                   &        &$^5D_2$=0.0317    &                  &                   &        &$^3D_1$=0.0122    & \\
                  & $D_sD_s$          & 0.13   &$^1D_2$=0.0139    &                  & $D_s^*D_s^*$      & 0.13   &$^5D_1$=0.0035    & \\
                  & $D_s^*D_s$        & 0.10    &$^3D_2$=0.0192    &                  & $DD_0$            & 0.03   &$^1P_1$=-0.0046   & \\
\hline
\end{tabular}
\end{table*}

\begin{table*}
\caption{\label{Tab1} Open-charm strong decay properties for the $1D$ and $2D$ charmonium states in the LP and SP models. }
\begin{tabular}{c   | c  c  c  c  c  c  c  l  c}\hline\hline
\multirow{2}{*}{State} & \multirow{2}{*}{Mode}& \multicolumn{4}{c}{\underline{~~~~~~~~~~~~~~~~~~~~~~~~~~~~LP model~~~~~~~~~~~~~~~~~~~~~~~~~~~~}} &\multicolumn{2}{c}{\underline{~~~~~SP model~~~~~}}&\underline{ Amps. (GeV$^{-1/2}$)} &\underline{ Mass (MeV)} \\
&    &      $\Gamma_{th}$ (MeV)   & Br (\%) & $\Gamma_{th}$ (MeV)~\cite{Barnes:2005pb}     & Br (\%)~\cite{Barnes:2005pb}   & $\Gamma_{th}$ (MeV)    & Br (\%) & ~~~~~~~~~~~LP/SP & LP/SP   \\ \hline
$\psi_{3}(1D)$  & $D^0\bar{D}^0$    & $0.52$ &59  & ...  & ... & $0.53$   &60   &$^1F_3$=0.015/-0.0153&$3811/3808$          \\
                & $D^+D^-$          & $0.36$ &41  & ...  & ... & $0.36$   &40   &$^1F_3$=0.0127/-0.0129             \\
                & Total             & $0.88$ &100 &$0.5$ &100  & $0.89$   &100  &\\ \hline
$\psi_{3}(2D)$  & $DD$              & $7.2$  &11  & $24$ &16   & $5.7$    &9    &$^1F_3$=-0.0243/-0.0321&$4172/4112$   \\
                & $D^*D$            & $29 $  &45  & $50$ &34   & $25$     &40   &$^3F_3$=-0.0756/-0.0752&               \\
                & $D^*D^*$          & $20 $  &31  & $67$ &45   & $28$     &44   &$^1F_3$=-0.0243/-0.0191&          \\
\hline\hline
\end{tabular}
\end{table*}

\begin{table*}
\caption{\label{tabdw} Open-charm strong decay properties for the $3D$ charmonium states in the LP and SP models. The widths listed in the brackets are calculated with the masses of observations. }
\begin{tabular}{c | c  c  c  c  c  l  c}\hline\hline
\multirow{2}{*}{State} & \multirow{2}{*}{Mode}& \multicolumn{2}{c}{\underline{~~~~~LP model~~~~~} }      &\multicolumn{2}{c}{\underline{~~~~~SP model~~~~~}} &\underline{ Amps. (GeV$^{-1/2}$)}&\underline{ Mass (MeV)}  \\
        &        &  $\Gamma_{th}$ (MeV)   & Br (\%) &$\Gamma_{th}$ (MeV)   & Br (\%) & ~~~~~~LP/SP & LP/SP \\
\hline
$\eta_{c2}(3D)$ & $D^*D$            & $16 $   &  24    & $13  $    &56    &$^3P_2$=-0.0177/-0.0178 &$4478/4336$  \\
                &                   &         &        &           &      &$^3F_2$=0.0415/0.0412   \\
                & $D^*D^*$          & $7.3$   &  11    & $5.2 $    &23    &$^3P_2$=-0.0006/0.005&  \\
                &                   &         &        &           &      &$^3F_2$=0.0324/0.0301   \\
                & ${D_s}^*D_s$      & $1.9$   &  2.8   & $1.6 $    &6.9   &$^3P_2$=-0.01/-0.0146& \\
                &                   &         &        &           &      &$^3F_2$=-0.0137/-0.0105  \\
\hline
\end{tabular}
\end{table*}

\begin{table*}
\caption{\label{tabdw2} Open-charm strong decay properties for the $3D$ charmonium states in the LP and SP models. The widths listed in the brackets are calculated with the masses of observations. }
\begin{tabular}{c | c  c  c  c  c  l  c}\hline\hline
\multirow{2}{*}{State} & \multirow{2}{*}{Mode}& \multicolumn{2}{c}{\underline{~~~~~LP model~~~~~} }      &\multicolumn{2}{c}{\underline{~~~~~SP model~~~~~}} &\underline{ Amps. (GeV$^{-1/2}$)}&\underline{ Mass (MeV)}  \\
        &        &  $\Gamma_{th}$ (MeV)   & Br (\%) &$\Gamma_{th}$ (MeV)   & Br (\%) &~~~~~~ LP/SP & LP/SP \\
\hline
$\psi_{2}(3D)$  & $D^*D$            & $13  $  & 21     & $11 $    & 42    &$^3P_2$=-0.0208/-0.0215& $4478/4337$ \\
                &                   &         &        &          &       &$^3F_2$=-0.0343/-0.0339&  \\
                & $D^*D^*$          & $9.8 $  & 16     & $6.9 $   & 26    &$^5P_2$=0.0003/0.0038 &  \\
                &                   &         &        &          &       &$^5F_2$=0.0376/0.0351  &   \\
                & ${D_s}^*D_s$      & $1.9 $  & 3.1    & $2.0 $   & 7.7   &$^3P_2$=-0.0129/-0.0182& \\
                &                   &         &        &          &       &$^3F_2$=0.0111/0.0087 &   \\
\hline\hline
\end{tabular}
\end{table*}

\subsection{Candidates of higher $c\bar{c}$ states with $J^{PC}=1^{++}$}

The higher $c\bar{c}$ states with $J^{PC}=1^{++}$, such as
$\chi_{c1}(2P)$, $\chi_{c1}(3P)$ and $\chi_{c1}(4P)$, are still not established.
During the past decade, several $J^{PC}=1^{++}$ charmonium-like states,
$X(3872)$, $X(4140)$ and $X(4274)$, have been observed
in experiments. They might be good candidates of the missing
$J^{PC}=1^{++}$ $c\bar{c}$ states, but associated by non-trivial dynamics.

\subsubsection{$X(3872)$}

The $X(3872)$ resonance has the same quantum numbers as $\chi_{c1}(2P)$ (i.e., $J^{PC}=1^{++}$)
but with a much lighter mass than potential quark model predictions. Its mass is close to the $DD^*$
threshold and makes it an ideal candidate for the $DD^*$ hadronic molecule. Various scenarios have been
discussed in the literature for which recent reviews can be found in Refs.~\cite{Chen:2016qju,Olsen:2014qna,
Guo:2017jvc,Olsen:2017bmm,Lebed:2016hpi}. It is now broadly accepted that the $X(3872)$ has both a
long-ranged molecular wave function and a short-ranged compact $c\bar{c}$ component~\cite{Chen:2013upa,Ferretti:2014xqa,Ferretti:2013faa,Ferretti:2018tco,Ferretti:2013vua}.
It can be viewed as the mixture of the $D\bar{D}^{*}+c.c.$ molecule and $\chi_{c1}(2P)$ combined by a
unitarized strong $S$-wave interaction ~\cite{Chen:2013upa} (see e.g. Ref.~\cite{Guo:2017jvc} for a detailed review).
According to other authors the $X(3872)$ can be interpreted as as $c\bar{c}$ core plus continuum
(meson-meson) components, like $D\bar{D}$, $D\bar{D}^*$, and so on as in
Refs.~\cite{Ferretti:2014xqa,Ferretti:2013faa,Ferretti:2018tco,Ferretti:2013vua}.

\begin{figure}[h]
\begin{center}
\centering  \epsfxsize=8.5 cm \epsfbox{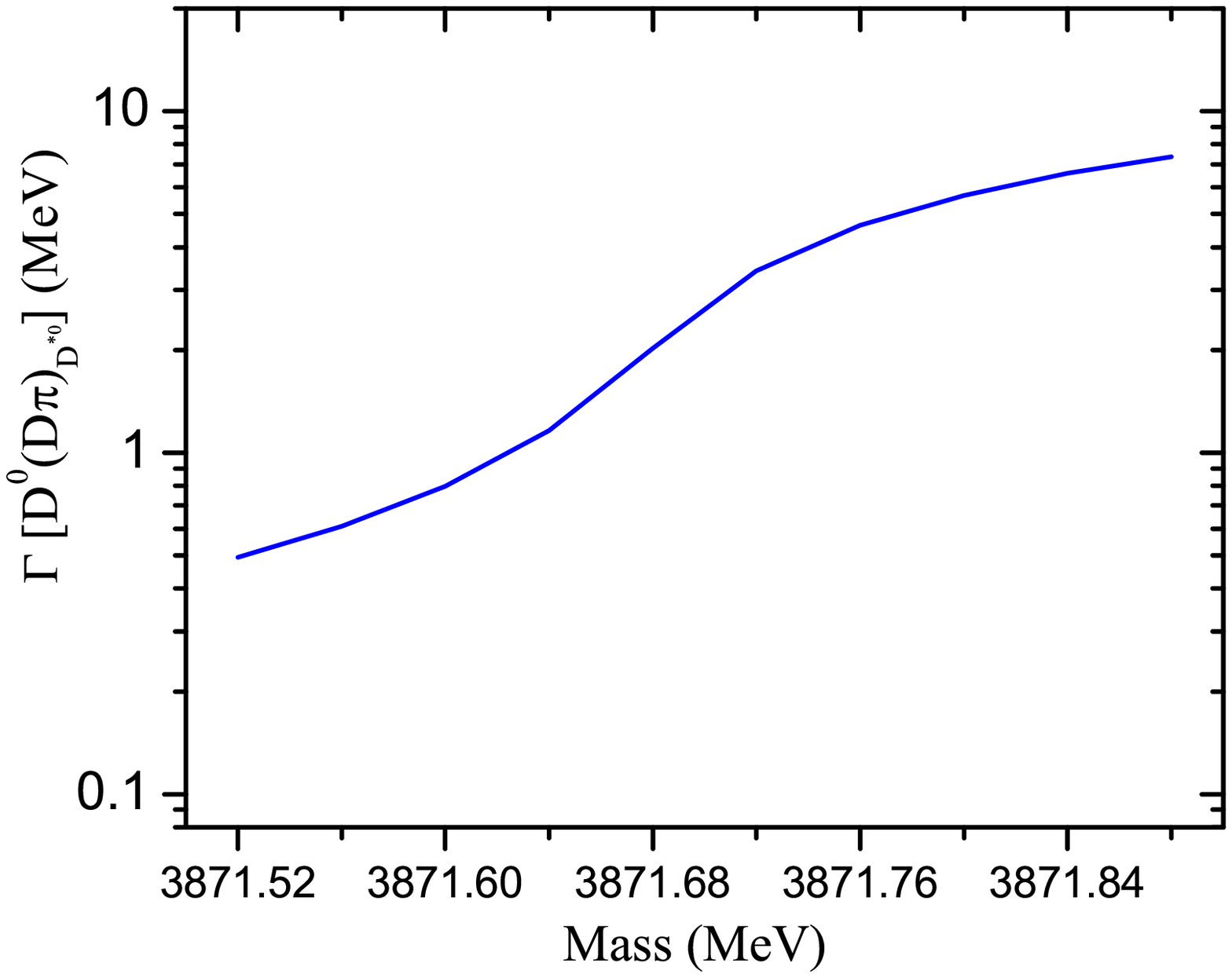} %\epsfxsize=5.0 cm \epsfbox{1DXB.eps}
\vspace{-0.3 cm} \caption{The quasi-two-body decay width of $X(3872)\to D^0(D\pi)_{D^{*0}}$ as a function of mass
by assigning $X(3872)$ as the $\chi_{c1}(2P)$ charmonium state.} \label{fig2}
\end{center}
\end{figure}

We do not expect that the LP and SP model can explain the observed properties of $X(3872)$ by
treating it as the $\chi_{c1}(2P)$ state. But as a test of the potential model calculations we
consider $X(3872)$ as the $\chi_{c1}(2P)$ state and calculate its strong decays into $D^0D^{*0}$.
It shows that a strong coupling can be extracted and the dominance of the partial widths of
$X(3872)\to D^0D^{*0}$  (if the input mass is higher than the threshold)
is consistent with the experimental observations.
We also predict the width of $X(3872)$ by using its physical mass $3871.69\pm 0.17$ MeV~\cite{Olive:2016xmw}
in the $^3P_0$ model, which gives a decay width of $\Gamma<7$ MeV.
It turns out to be that the decay width is very sensitive to the mass of $X(3872)$.
As the physical mass of $X(3872)$ is very close to the $D^0\bar{D}^{*0}$ threshold,
the line shape of $D^{*0}$ should be taken into account.
This effect can be included by using the quasi-two-body decay formula \cite{Ferretti:2014xqa,Kokoski:1985is,Capstick:1993kb,Roberts:1997kq,Segovia:2009zz}
\begin{align}
    \Gamma [D^0(D\pi)_{D^{*0}}] &=2\int_{0}^{k_{max}}dkk^{2} \sum_{J,L}| \mathcal{M}^{JL}(X(3872)\to D^0\bar{D}^{0*})|^{2} \nonumber \\
    & \times \frac{\Gamma_{\bar{D}^{0*}\to\bar{D}^{0}\pi^{0}}(k)}{|M_{X(3872)}-E_{\bar{D}^{0*}}(k)-E_{D^{0}}(k)|^{2}+\frac{1}{4}\Gamma_{\bar{D}^{0*}}^{2}},
    \label{QuasiTB}
\end{align}
where $\mathcal{M}^{JL}(X(3872)\to D^0\bar{D}^{0*})$ is the $^3P_0$ amplitude defined in Eq.(\ref{eq4}),
$\Gamma_{\bar{D}^{0*}\to\bar{D}^{0}\pi^{0}}(k)$ is the energy-dependent
decay width of the unstable meson $\bar{D}^{*0}$ worked out within the chiral quark model~\cite{Zhong:2008kd},
and $\Gamma_{\bar{D}^{*0}}$ is the total decay width of $\bar{D}^{*0}$. According to PDG, we have $\Gamma_{\bar{D}^{*0}}=\Gamma_{\bar{D}^{0*}\to\bar{D}^{0}\pi^{0}}/0.65$~\cite{Olive:2016xmw}.
As the unstable meson $D^{*0}$ can slightly off shell, its momentum $k$ can range from 0 to
$k_{max}=\frac{\sqrt{M_{X(3872)}^2-(M_{\bar{D}^0}+M_{\pi^0}+M_{D^0})^2}\sqrt{M_{X(3872)}^{2}-(M_{\bar{D}^{0}}+M_{\pi^{0}}-M_{D^{0}})^{2}}}{2M_{X(3872)}} $. Using Eq.(\ref{QuasiTB}), we plot the decay width of $X(3872)$ as a function of its mass within one sigma range in Fig. \ref{fig2}.
We can see that there is no zero decay width when $X(3872)$ under $D^0D^{0*}$ threshold as $D^{0*}$ can be off shell. When the $D^0 D^{0*}$ threshold opens, the decay width of $X(3872)$ increases rapidly.
Finally, the decay width of $X(3872)$ is predicted to be $\Gamma[X(3872)]\simeq 2 $ MeV
when we take the world average mass $m_{X(3872)} = 3871.69$ MeV from the PDG~\cite{Olive:2016xmw}.
This result is consistent with the measured widths in experiments~\cite{Adachi:2008sua,Aubert:2007rva}.

We mention that the radiative
transitions of $X(3872)\to \psi(2S) \gamma$ and $J/\psi \gamma$ were studied in
Ref.~\cite{Deng:2016stx,Ferretti:2014xqa}, where the $X(3872)$ was also treated as the $\chi_{c1}(2P)$ state.
It also shows that the radiative decay properties are consistent with the observations from the BaBar~\cite{Aubert:2008ae} and
LHCb~\cite{Aaij:2014ala}. This feature was regarded as evidence for $X(3872)$ being the $\chi_{c1}(2P)$ state. However, as studied by Refs.~\cite{Mehen:2011ds,Guo:2014taa} the radiative decays of $X(3872)$ are shown to be driven by the short-ranged component of the wave function. Therefore, our results actually can be regarded as a support of such a view that the short-ranged wave function is from the $\chi_{c1}(2P)$ state instead of concluding that it is a $\chi_{c1}(2P)$ state. Since the physical state of $X(3872)$ is apparently different from a conventional charmonium state, and there have been tremendous works discussing its dynamics, we do not want to over-interpret it based on our approach.

\subsubsection{$X(4140)$ and $X(4274)$}

In the SP model, the  mass of $\chi_{c1}(3P)$ is predicted to be $\sim 4.19$ MeV. Thus, the charmonium-like state $X(4140)$ can be a candidate for
the $\chi_{c1}(3P)$ state. In this scenario its total width is found to be $\sim 14$ MeV, which is in agreement with the world average data $19^{+8}_{-7}$ MeV~\cite{Olive:2016xmw}. If $X(4140)$ corresponds to the $\chi_{c1}(3P)$ indeed, it may mainly decay into $DD^*$, $D^*D^*$ and $D_sD^*_s$ channels with comparable decay rates. The partial width ratios between these decay modes are predicted to be
\begin{equation}
\frac{\Gamma(DD^*)}{\Gamma(D^*D^*)}\simeq 0.69, ~ \frac{\Gamma(D_sD_s^*)}{\Gamma(D^*D^*)}\simeq 0.40,
\end{equation}
which can be tested in future experiment.

In contrast with the LP model calculations, the  mass of $\chi_{c1}(3P)$ is predicted to be
$\sim 4.28$ MeV which is close to $X(4274)$. Assigning $X(4274)$ as the $\chi_{c1}(3P)$ state,
we calculate its strong decays in the LP model. It shows that the LP model produces a narrow width
of about 21 MeV which is close to the lower limit of the measured width $56\pm 11^{+8}_{-11}$ MeV~\cite{Aaij:2016iza}.
The ratios between different partial widths, i.e. $DD^*$, $D^*D^*$ and $D_sD^*_s$, are
predicted to be
\begin{equation}
\frac{\Gamma(D^*D^*)}{\Gamma(DD^*)}\simeq 2\times 10^{-3}, ~ \frac{\Gamma(D_sD_s^*)}{\Gamma(DD^*)}\simeq 1.8 \ .
\end{equation}

It is interesting to note that the recent analysis of the $B^+\to J/\psi \phi K^+$
process~\cite{Liu:2016onn}, and the study of the masses of $cs\bar{cs}$ tetraquark
states~\cite{Lu:2016cwr} seem to favor that the $X(4274)$ may be the conventional $\chi_{c1}(3P)$ state.
If $X(4274)$ is assigned as the $\chi_{c1}(3P)$ state, then one
immediate question is how to understand $X(4140)$. In Ref.~\cite{Liu:2016onn} the analysis suggests that the $X(4140)$ structure
may not be a genuine resonance. There are possibilities for a non-resonance interpretation for $Y(4140)$, such as the $D_sD_s^*$ CUSP~\cite{Aaij:2016iza,Aaij:2016nsc}, or $D_sD_s^*$ rescatterings via the open-charmed meson loops~\cite{Liu:2016onn}.
To better understand $X(4140)$ and $X(4274)$, experimental studies of their open-charm decay modes,
i.e. $DD^*$, $D^*D^*$ and $D_sD^*_s$, are strongly recommended.

\subsection{Candidates of higher $c\bar{c}$ states with $J^{PC}=0^{++}$}

The higher $c\bar{c}$ states with $J^{PC}=0^{++}$, such as
$\chi_{c0}(2P)$, $\chi_{c0}(3P)$ and $\chi_{c0}(4P)$, have not yet been established.
Recently, several $J^{PC}=0^{++}$ charmonium-like states,
$X^*(3860)$ and $X(4500,4700)$ have been observed
at Belle~\cite{Chilikin:2017evr} and LHCb~\cite{Aaij:2016iza}, respectively.
They might be good candidates for these missing
$J^{PC}=0^{++}$ $c\bar{c}$ states.

\subsubsection{$X^*(3860)$}

The charmonium-like state $X^*(3860)$ observed in the $e^+e^-\to J/\psi D\bar{D}$
by the Belle Collaboration~\cite{Chilikin:2017evr} serves a good candidate for the $\chi_{c0}(2P)$ state. Its measured mass and width are $3862^{+26+40}_{-32-13}$ MeV and $201^{+154+88}_{-67-82}$ MeV, respectively, and fits the expectation of the $\chi_{c0}(2P)$ state
predicted in the potential models.

Considering $X^*(3860)$ as the $\chi_{c0}(2P)$ state we study its strong decays into $DD$ in the LP and SP models.
Our results are listed in Table~\ref{tab3}. Both models give a similar value of $\Gamma\simeq  22\sim 28$ MeV, which is rather narrow. Our results are consistent with that from Ref.~\cite{Barnes:2005pb}. Note that the present experimental width is not well determined. In contrast, results from different theoretical calculations are still controversial. For instance, the  recent analysis of Ref.~\cite{Zhou:2017dwj} found a small width about 11 MeV for the $\chi_{c0}(2P)$ state~\cite{Zhou:2017dwj}, while the analysis of Ref.~\cite{Yu:2017bsj} in a $^3P_0$ model obtained a large width of $110\sim 180$ MeV. This suggests that further precise measurement of the $X^*(3860)$ and more theoretical studies of the $\chi_{c0}(2P)$ state are necessary. We mention that tetraquark interpretations were also proposed for  the nature of
$X^*(3860)$ in the literature~\cite{Wang:2017lbl,Chen:2017dpy}.

\subsubsection{$X(4500/4700)$}

Recently, two $J^{PC}=0^{++}$ charmonium-like state $X(4500)$
and $X(4700)$ were observed in the $J/\psi \phi$ invariant mass distributions in $B^+\to J/\psi \phi K^+$ at LHCb~\cite{Aaij:2016iza,Aaij:2016nsc}.
The measured widths of $X(4500)$ and $X(4700)$ are
$92\pm 21^{+21}_{-20}$ MeV and $120\pm 31^{+42}_{-33}$ MeV, respectively.
To understand the nature of these two structures,
different interpretations have been proposed in the literature~\cite{Lu:2016cwr,Chen:2016oma,Wu:2016gas,Wang:2016ujn,Wang:2016gxp}. In this work we consider the possible assignments of these two states as conventional charmonium states.

In the LP model the predicted mass of $\chi_{c0}(4P)$, $M\simeq 4544$ MeV,
is close to the mass of $X(4500)$, $4506\pm 11^{+12}_{-15}$ MeV.
Considering the $X(4500)$ as the $\chi_{c0}(4P)$ state, we study its strong decay
properties and the results are listed in Table~\ref{amp4}.
It is shows that the calculated total
width, $\Gamma\sim 50$ MeV, is compatible with the measured value
of $X(4500)$ within its uncertainties. In Ref.~\cite{Ortega:2016hde}, the $X(4500)$ was also
considered as the conventional $\chi_{c0}(4P)$ state in the coupled-channel approach.
If $X(4500)$ is the $\chi_{c0}(4P)$ state indeed,
it may dominantly decay into the $DD_1$, $DD_1'$, $D^*D_0$ and $D^*D_2$ channels, and
the branching fractions are about $\mathcal{O}(10\%)$. In contrast, the decay rate into $DD$ channel is relatively small, which is
is predicted to be $Br[X(4500)\to DD]\sim 10^{-3}$. Such a small branching ratio may be difficult to observe
in experiment.

In the SP model the predicted mass of $\chi_{c0}(5P)$ is $4537$ MeV,
which also makes $X(4500)$ a good candidate. Assigning $X(4500)$
as the $\chi_{c0}(5P)$ state, the strong decay properties of $X(4500)$ are studied and the results are listed in Table~\ref{amp6}. It shows that the $\chi_{c0}(5P)$ state should be very narrow in the SP model with a width
of $\Gamma\simeq 15$ MeV. Its dominant decay modes are the $DD_1$ and $DD_1'$ channels. One notices that the total width in the SP model is about a factor of 4 smaller than the observed width of $X(4500)$.

In the SP model the predicted mass of $\chi_{c0}(6P)$ is about $4669$ MeV,
which is very close to that of $X(4700)$. Considering $X(4700)$ as
the $\chi_{c0}(6P)$ state, we find that the predicted width is about 16 MeV (see Table~\ref{amp8}),
which is too narrow to be comparable with the observed width $120\pm 31^{+42}_{-33}$ MeV
of $X(4700)$. Although the mass seems to fit the experimental observation, the significant discrepancy in the width raises questions on the structure of $X(4700)$. In Ref.~\cite{Liu:2016onn}, the $X(4700)$ is explained as the $\psi'\phi$ rescattering via the $\psi'K_1$ loops.
The $X(4700)$ resonance might be a candidate of charmonium hybrid, for its mass is close to the charmonium hybrids with $J^{PC}=0^{++}$ predicted in Ref.~\cite{Guo:2008yz}.

In brief, we find that as the $\chi_{c0}(4P)$ state the mass and width of $X(4500)$ can be understood in the LP model.
In contrast, although the masses of both $X(4500)$ and $X(4700)$ can be described by the SP model,
their widths appear to be difficult to understand.

\subsection{Candidates of higher $c\bar{c}$ states with $J^{PC}=2^{++}$}

The $X(4350)$ found by Belle~\cite{Shen:2009vs}
in the $\phi J/\psi$ mass spectrum has been a good candidate for the $\chi_{c2}(3P)$ state with $J^{PC}=2^{++}$. The extracted mass and width are
$4350.6^{+4.6}_{-5.1}\pm 0.7$ MeV and $13^{+18}_{-9}\pm 4$ MeV, respectively. In Ref.~\cite{Liu:2009fe} such a possible assignment was also considered.

In the LP model the calculated mass of
$\chi_{c2}(3P)$ is about $4310$ MeV, which is very close to the mass of $X(4350)$.
Assigning the $X(4350)$ as the $\chi_{c2}(3P)$, we study its strong decays and the results are listed
in Table.~\ref{tab3}. Its width is found to be about $90$ MeV,
which is much larger than the observed value. This indeed raises questions on such an assignment. Note that the present experimental information is still very rough. Therefore, future experimental search for its decays into $DD^*$,  $DD$,
$D^*D^*$ and $D_s^*D_s^*$ are strongly recommended.

\subsection{Candidates of higher $c\bar{c}$ states with $J^{PC}=0^{-+}$}
%\subsection{$X(3940)$ and the $\eta_c(3S)$ state }

The $X(3940)$ was first observed by the Belle Collaboration in
$e^+e^-\to J/\psi+X$~\cite{Abe:2007jna}. This state is also established in
the invariant mass spectrum of $D^*\bar{D}$ in
$e^+e^-\to J/\psi D^*\bar{D}$ process~\cite{Abe:2007sya}. The updated mass and width
of $X(3940)$ are $M=3942^{+7}_{-6}\pm 6$ MeV and $\Gamma=37^{+26}_{-15}\pm 8$ MeV, respectively.
Its decay into $DD^*$ but not $DD$ suggests that
it has unnatural parity. The most likely interpretation of $X(3940)$ is that it
is the $\eta_c(3S)$ state with $J^{PC}=0^{-+}$~\cite{Eichten:2007qx}, although the predicted
mass in potential models appears to be higher than the observations.

Considering $X(3940)$ as the $\eta_c(3S)$ state, we analyze its strong decays in both the
LP and SP models and the results are listed in Table~\ref{tabs}. Due to the limited phase space, it shows that the $DD^*$ channel is the only open-charm decay for $X(3940)$.
Furthermore, in the SP model the decay width is predicted to be about $40$ MeV, which is in good
agreement with the measurements. We also mention that the width is consistent with the calculation result from the
Bethe-Salpeter method~\cite{Wang:2016mqb}. So far, the $X(3940)$ turns out to be a good candidate for $\eta_c(3S)$
according to its strong decay properties.

\section{Summary}\label{sum}

In this work we carry out a systematical study of the open-charm strong decays of the higher charmonium states up to
the mass of the $6P$ multiplet within the $^3P_0$ model. The wave functions of the initial charmonium
states are adopted from the calculations by the LP and SP models in our previous work.
Several key results from this study can be learned here:

\begin{itemize}

\item The decay widths for the well-established
charmonium states $\psi(3770)$, $\psi(4040)$, $\psi(4160)$
and $\chi_{c2}(2P)$ can be reasonably described in the LP and SP models, although for $\psi(4040)$ and $\psi(4160)$
some partial width ratios between the open-charm decay modes appear to have large discrepancies
with the data.

\item In the LP model, the $\psi(4415)$ favors the $\psi(4S)$ assignment,
while the possibility of $\psi(4415)$ as the $\psi_1(3D)$ state cannot be excluded.
There may exist two highly overlap $J^{PC}=1^{--}$ charmonium states around 4.4 GeV.

\item In the LP model, the charmonium-like states $X(4500)$ and $X(4660)$ could be assigned as
the $\chi_{c0}(4P)$ and $\psi(5S)$ states, respectively.

\item The newly observed state $X^*(3860)$ seems to be too broad if it is classified as the
$\chi_{c0}(2P)$ state. Our calculation shows that the
$\chi_{c0}(2P)$ state should be a narrow state with $\Gamma\simeq 30$ MeV.

\item The $X(3940)$ favors the assignment as the $\eta_c(3S)$ state,
although the predicted mass in the potential models is somehow higher than the experimental data.

\item  The $X(4140)$ resonance may be a good candidate of the
$\chi_{c1}(3P)$ state in the SP model, while in the LP model $X(4274)$ seems to favor the $\chi_{c1}(3P)$ state. However, since it is difficult to accommodate these two states in the same model, future studies of these two states are needed for a better understanding of their nature.

\item  The vector charmonium-like states, $Y(4230/4260)$ and $Y(4360)$, and the scalar charmonium state $X(4700)$
cannot be accommodated by the conventional charmonium spectrum.

\end{itemize}

%Finally, it should be mentioned that in the SP model the
%predicted masses for the charmonium states $\psi(4S)$, $\psi(5S)$,
%$\psi_1(3D)$, and $\chi_{c0}(6P)$ are comparable to the masses
%of $X(4260)$, $\psi(4415)$, $X(4360)$, and $X(4700)$, respectively,
%however, their widths predicted in theory are inconsistent with the observations.

In brief, we show that a systematic study of the charmonium spectrum in the LP and SP models is useful for identifying unusual features arising from some of those higher charmonium-like states recently observed in experiment. We anticipate that future experimental measurements of some of those open-charm decay channels should be helpful for pinning down both conventional and unconventional charmonium states and provide more insights into the underlying dynamics in the charmonium mass regime.

%there still exist many puzzles in our understanding
%of the nature of the higher charmonium states.
%To clarify these puzzles, the dominant open-charm decay
%channels of the higher charmonium states are suggested to be observed in forthcoming experiments.
%We also hope that our predictions could be helpful for future experimental
%search for some of those excited charmonium states.

\section*{  Acknowledgements }

This work is supported, in part, by the National Natural Science Foundation of China under Grants
No. 11775078, No. 11405053, No. 11705056,
No. 11425525, No. 11521505, and No. 11261130311; by DFG and NSFC through
funds provided to the Sino-German CRC 110 ``Symmetries and the Emergence of
Structure in QCD'' (NSFC Grant No. 11261130311),
and National Key Basic Research Program of China under Contract No. 2015CB856700.

\end{document}